\documentclass[%
 reprint,
nofootinbib,
 amsmath,amssymb,
 aps,
]{revtex4-2}

\usepackage{xcolor,graphicx}
\usepackage{dcolumn}
\usepackage{bm}
\usepackage{subcaption}
\usepackage[colorlinks,linkcolor=blue]{hyperref}

\def\al{\alpha}
\def\be{\beta}
\def\ga{\gamma}
\def\de{\delta}

\def\ka{\kappa}
\def\la{\lambda}

\def\Ga{\Gamma}

\def\cL{{\mathcal L}}
\def\cR{{\mathcal R}}
\def\cD{{\mathcal D}}

\def\lie{{\cL_{\bf n}}}

\def\mn{{\mu\nu}}
\def\ab{{\al\be}}

\def\pt#1{\phantom{#1}}

\def\3g#1#2#3{^{(3)}\Ga^{#1}_{\pt{#1}#2#3}}

\newcommand{\beq}{\begin{equation}}
\newcommand{\eeq}{\end{equation}}
\newcommand{\bea}{\begin{eqnarray}}
\newcommand{\eea}{\end{eqnarray}}

\begin{document}

\preprint{number}

\title{Explicit spacetime-symmetry breaking and the dynamics of primordial fields}

\author{Nils A. Nilsson}
 \email{nilssonnilsalbin@gmail.com}
\affiliation{%
Center for Quantum Spacetime, Sogang University, Seoul, 121-742, Korea
}%
\affiliation{%
SYRTE, Observatoire de Paris-PSL, Sorbonne Universit\'e, CNRS UMR8630, LNE, 61 avenue de
l’Observatoire, 75014 Paris, France
}%

\date{\today}

\begin{abstract}
We study the effects of explicit spacetime-symmetry breaking on primordial tensor fluctuations using an effective-field theory for Lorentz/CPT violation. We find that the graviton is still massless, but that the propagation speed of tensor modes is modified, and we obtain a constraint on the coefficient determining the symmetry breaking on the order of $10^{-15}$ from the recent measurements of the speed of gravity. Due to the symmetry breaking, the de-Sitter phase is modified, and during this inflationary epoch, the power spectrum assumes a slow oscillation around the general-relativity limit; further, we find that the primordial tensor power spectrum retains its scale invariance, but that the amplitude is modified. We also find that the modes which become subhorizon during radiation domination acquire a phase shift proportional to the coefficient for Lorentz violation.

\end{abstract}

\maketitle


\section{\label{sec:intro}Introduction}
The recent detection of gravitational waves (GW's) by the LIGO and Virgo collaboration not only confirmed yet another prediction of general relativity, but opened a completely new window to the universe. GW's can now be used to study black holes and neutron stars unlike any other probe, and can also put bounds on the mass of the graviton and the ``speed of gravity`` when combined with electromagnetic observations. Different mechanisms for generating GW's lie in the early universe, where a number of different processes may create primordial GW's (PGW's); indeed, the existence of PGW's is an important prediction of the inflationary model of the early universe~\cite{Grishchuk:1974ny,Starobinsky:1979ty}. Even prior to inflation, the existence of PGW's have been shown to appear in Big Bounce models~\cite{Malkiewicz:2020fvy}. Post-inflation, early-universe mechanisms such as BBN nucleosynthesis, phase transitions~\cite{Kosowsky:1991ua,Kamionkowski:1993fg}, cosmic pre/reheating~\cite{Easther:2006gt,Khlebnikov:1997di}, and cosmic defects~\cite{Vachaspati:1984gt,Figueroa:2012kw} can all give rise to PGW's, as can other scenarios such as the merging of primordial black holes~\cite{Garcia-Bellido:1996mdl,Dong:2015yjs} or the presence of extra fields and broken spacetime symmetries in the early universe. 

Most early-universe mechanisms are expected to produce exceedingly weak PGW's; for example, in the slow-roll scenario, the energy density of PGW's are expected to be approximately $\Omega_{\rm GW}(f) \approx 10^{-15}$ for many orders of magnitude in the frequency $f$. Such a low value is unlikely to be accessible by ground-based observatories, but may be observable by future space-based detectors, for example ALIA, BBO, and DECIGO~\cite{PhysRevD.72.083005,Kawamura:2011zz}. It may be possible, however, to generate observable signals within models beyond standard inflation and cosmology~\cite{PhysRevD.73.063008,Bernal:2019lpc}. Any detection of a stochastic background of primordial gravitational waves would have profound implications for our understanding of the early universe, as they may give insight into processes at very early times, probing further than the recombination era (redshift $z\approx 1100$) and the release of the CMB~\cite{Grishchuk:1974ny,Fabbri:1983us,White:1992fj,Sahni:1990tx}. 

Since the early universe lends itself well to the study of fundamental physics, it is also a promising testing ground for spacetime symmetries; indeed, it has been suggested that local Lorentz invariance may no longer be an exact symmetry in a fundamental, unified theory of physics~\cite{KostelStuart1989,Ellis:1999jf}. In fact, several approaches to quantum gravity predict or allow for the breaking of exact Lorentz/CPT symmetry~\cite{Ellis:1999jf,Gambini:1998it,Alfaro:1999wd,Alfaro:2001rb}.
Since local Lorentz invariance is a fundamental ingredient of both general relativity and particle physics, stating that any local experiment is independent of both orientations and velocities, performing precision tests of Lorentz symmetry itself is an excellent way to test for new physics. 

These above considerations has lead to a large research effort in the testing of spacetime symmetries using a generalised effective-field theory framework, making it possible to combining results from very different experiments, even across Standard Model sectors~\cite{PhysRevD.69.105009,PhysRevD.58.116002,PhysRevD.55.6760}. For linearised gravity, limits exist from Solar-system tests~\cite{Iorio:2012gr,doi:10.1142/9789814623995_0440,PhysRevD.94.125030}, short-range gravity~\cite{PhysRevD.91.092003,PhysRevLett.117.071102,PhysRevLett.122.011102,Bailey:2022wuv}, pulsar tests~\cite{PhysRevD.98.084049,PhysRevLett.112.111103}, gravitational waves~\cite{ONeal-Ault:2021uwu,Wang:2021ctl,PhysRevD.101.104019,PhysRevD.102.024028}, and many more. The results in the exact (non-linearised) regime are less exhaustive at present, and have only just begun to be explored; the effects of exact spacetime-symmetry breaking have been studied in for example inflation~\cite{Bonder:2017dpb}, black holes~\cite{Bonder:2020fpn}, cosmology~\cite{ONeal-Ault:2020ebv}, and others~\cite{Bailey:2016ezm,Bonder:2015maa,Bailey:2019rjj,Reyes:2021cpx}. A complete list of all available Lorentz and CPT violation constraints within this framework can be found in~\cite{Kostelecky:2008ts}. 

There is a clear distinction between spontaneous and explicit symmetry breaking: with spontaneous breaking, all the canonical relations, such as the Noether identities and Euler-Lagrange equations still hold. In contrast, in the case of explicit breaking, the terms breaking the symmetry appear in the Lagrangian as nondynamical background tensors which have preferred spacetime directions. These tensors do not obey equations of motion, and the notion of being ``on shell'' cannot be applied to them. This creates conflicts with the Bianchi identities and energy momentum conservation, often imposing such severe constraints that the theory cannot survive; however, there are scenarios in which explicit spacetime-symmetry breaking can be made to fit with all the requirements imposed by Einstein's equations and the notions of Riemannian geometry~\cite{Bluhm:2019ato}. It also bears mentioning that when using a Stuckelberg approach to explicit breaking, the extra degrees of freedom takes on the same form as the massless Nambu-Goldstone excitations present in the case of spontaneous symmetry breaking, as noted in \cite{Bluhm:2019ato}.

The purpose of the present paper is to study the consequences of explicit spacetime-symmetry breaking on the dynamics of primordial gravitational waves. By building upon previously obtained results~\cite{ONeal-Ault:2020ebv}, we explore a simple case of explicit spacetime-symmetry breaking in the context of the early Universe.

This paper is composed as follows: In Section~\ref{sec:theory} we set up the model and present cosmological solutions and conservation equations at the background level. In Section~\ref{sec:perts}, we carry out the perturbative analysis and find the equations of motion for the tensor modes, as well as solve the Mukhanov-Sasaki equation and find the expression for the power spectrum. We discuss our results and put them into context in Section~\ref{sec:disc}.
Throughout this paper, we use units in which $c=\hbar=1$, and the metric signature $(-,+,+,+)$. We use Greek letters $\mu,\nu,\alpha,\hdots$ as spacetime indices and mid-alphabet Latin letters $i,j,k,\hdots$ as spatial indices.

\section{Theoretical setup\label{sec:theory}}
Restricting to mass-dimension $d\leq 4$, we write the Lagrangian as
\begin{equation}
\mathcal{L} \sim R+(k_{\rm R})^{\alpha\beta\mu\nu}R_{\alpha\beta\mu\nu},
\end{equation}
where $R_{\alpha\beta\mu\nu}$ is the Riemann tensor and $(k_{\rm R})^{\alpha\beta\mu\nu}$ are the coefficients for Lorentz violation. $(k_{\rm R})^{\alpha\beta\mu\nu}$ can be decomposed into a scalar, trace-free, and trace part: $k_{\rm R} \rightarrow -uR + s_T^{\mu\nu}R_{\mu\nu}^{(\rm T)}+t^{\alpha\beta\mu\nu}W_{\alpha\beta\mu\nu}$. Here $R^{(\rm T)}_{\mu\nu}$ is the trace-free Ricci tensor, and $W_{\alpha\beta\mu\nu}$ is the Weyl tensor. The coefficients $u$, $s_T^{\mu\nu}$, $t^{\alpha\beta\mu\nu}$ break local \emph{particle} Lorentz and diffeomorphism symmetry, and do not carry dynamics in the explicit-breaking case (note that $u$ can be absorbed (if we assume it is constant) by a redefinition of the gravitational constant, and is thus unobservable. We may also move it to the matter sector). Therefore, $u$, $s^{\mu\nu}$, $t^{\alpha\beta\mu\nu}$ are \emph{not} solutions to the equations of motion and instead act as objects with prior geometry. This is in contrast to the case of spontaneous symmetry breaking, when the coefficients for Lorentz violation are vacuum expectation values of underlying dynamical fields obeying equations of motion, and are accompanied by both Nambu-Goldstone modes and massive modes; in this case, the action is invariant under particle diffeomorphisms~\cite{Bluhm:2021lzf,Bluhm:2019ato}. 

Since the $u$-term in the above decomposition can always be reincorporated as the trace-part of the coefficient $s^{\mu\nu}_T$, we rewrite the coefficient $k_{\rm R} = s_{\mu\nu}R^{\mu\nu}+t^{\alpha\beta\mu\nu}W_{\alpha\beta\mu\nu}$, where $s_{\mu\nu}$ is no longer trace-free. This is the object we will use in the entirety of the paper.

We take as our starting point the decomposition of a 4-dimensional globally hyperbolic manifold $\mathcal{M}$ into constant-time spatial hypersurfaces $\Sigma_t$ endowed with a timelike normal vector $n_\mu n^\mu = -1$. Using the Arnowitt-Deser-Misner description \cite{Arnowitt:1962hi}, the spatial metric takes the form $\gamma^{\mu\nu} = g^{\mu\nu}+n^\mu n^\nu$ (which also acts as the projection operator from $\mathcal{M} \rightarrow \Sigma_t$), after which the full metric can be written as
\begin{equation}\label{eq:ADMmetric}
    ds^2 = -(\alpha^2-\beta^i\beta_j)dt^2 + 2\beta_j dtd^j + \gamma_{ij}dx^idx^j,
\end{equation}
where $\alpha$ is the lapse function, $\beta_j$ is the shift vector, and $\gamma_{ij}$ is the spatial metric.
Considering now the case where only the coefficient $s_{\mu\nu}$ is non-zero, we arrive at the following ADM Lagrange density \cite{ONeal-Ault:2021uwu}
\bea\label{eq:mainLagr}
\cL &=& \cL_{\rm GR} + \sqrt{-g} [ s_\mn \cR^\mn  
- n^\al n^\be s_\ab (K^\mn K_\mn - K^2 )
\nonumber\\
&& \pt{-g} 
+2 s_\ab K^{\al\de} K^\be_{\pt{\be}\de} 
+K^\mn \lie s_\mn 
-K \lie (n^\mu n^\nu s_\mn) 
\nonumber\\
&& \pt{-g} 
+ 2 K \left( s_\mn n^\mu a^\nu + \cD_\la (s_\mn n^\mu \ga^{\nu\la})\right) 
\nonumber\\
&&\pt{-g} 
- 2 K^\la_{\pt{\la}\ka} \cD_\la (s_\mn n^\mu \ga^{\nu\ka} )
\nonumber\\
&&\pt{-g} 
+ a_\ka \cD_\la (s_\mn \ga^{\mu\la} \ga^{\nu\ka})
- a^\la \cD_\la ( s_\mn n^\mu n^\nu )],
\eea
where $\mathcal{R}^{\mu\nu}$ is the three-dimensional Ricci tensor, $\mathcal{D}$ is the covariant derivative on the spatial hypersurface, and $a_\mu = n^\nu\nabla_\nu n_\mu$ is the ADM acceleration. The extrinsic curvature is defined as $K_{\mu\nu}=-\nabla_\mu n_\nu-n_\mu a_\nu$, but can also be expressed through the Lie derivative of the spatial metric along the normal vector as $K_{\mu\nu} = -\tfrac{1}{2}\mathcal{L}_n \gamma_{\mu\nu}$. The GR part is obtained as
\begin{equation}
    \cL_{\rm GR} = \sqrt{-g}[\mathcal{R}+K^{\alpha\beta}K_{\alpha\beta}-K^2-2\nabla_\alpha(n^\alpha K+a^\alpha)],
\end{equation}
where the last term is taken to vanish on the boundary.

We continue along the lines of \cite{ONeal-Ault:2021uwu} and introduce the (significant) simplification where only \emph{one} component of the tensor $s_{\mu\nu}$ is non-vanishing, i.e.
\begin{equation}
    s_{\mu\nu} = \begin{pmatrix} s_{00} & 0 & 0 & 0 \\
    0 & 0 & 0 & 0 \\
    0 & 0 & 0 & 0 \\
    0 & 0 & 0 & 0 \\
    \end{pmatrix};
\end{equation}
using this, the Lagrangian (\ref{eq:mainLagr}) reduces to
\begin{align}\label{eq:Lagrred}
   \nonumber \mathcal{L} =& \frac{\alpha\sqrt{\gamma}}{2\kappa}\Big[\mathcal{R}+(1-\frac{s_{00}}{\alpha^2})\left(K_{ij}K^{ij}-K^2\right) +\frac{2}{\alpha^2}s_{00}a^ia_i\\ & \nonumber K\left(\frac{2}{\alpha^4}s_{00}(\dot{\alpha}-\alpha\beta^ia_i)-\frac{1}{\alpha^3}(\dot{s}_{00}-\beta^i\partial_is_{00})\right) \\& -\frac{1}{\alpha^2}a^i\partial_is_{00}\Big].
\end{align}
We will use the Friedmann-Lemaitre-Robertson-Walker (FLRW) metric, which reads
\begin{equation}\label{eq:flrw}
    ds^2 = -dt^2 + a^2(t)\left(\frac{dr^2}{1-kr^2}+r^2d\theta^2 + r^2\sin{\theta}^2d\phi^2\right),
\end{equation}
where $a(t)$ is the cosmic scale factor and $k$ denotes spatial curvature, with $k=\{+1,0,-1\}$ represents a closed, flat, and open universe, respectively. We use this metric to derive all background quantities, but we later reduce the equations to the flat ($k=0$) case.

We now proceed to restrict the coordinate dependence of $s_{00}$ by requiring that it is a constant within the constant-time spatial hypersurfaces $\Sigma_t$, $\partial_i s_{00} = 0$. This partial derivative is related to the contracted Bianchi identities for $s_{\mu\nu}$ \cite{PhysRevD.69.105009}
\begin{equation}\label{eq:bianchi}
    \nabla_\mu(T_s)^\mu_{~\nu} = \tfrac{1}{2}R^{\mu\lambda}\nabla_\nu s_{\mu\lambda}-\nabla_{\mu}(R^{\mu\lambda}s_{\lambda\nu}),
\end{equation}
where $(T_s)^\mu_{~\nu}$ is the stress-energy tensor for $s_{\mu\nu}$, which is related to the stress-energy tensor for matter $(T_M)^\mu_{~\nu}$ as $\nabla_\mu(T_s)^\mu_{~\nu} = -\kappa \nabla_\mu(T_M)^\mu_{~\nu}$. For matter, we add the usual perfect fluid for a homogeneous and isotropic Universe $(T_M)^\mu_{~\nu} = \text{diag}(-\rho,p,p,p)$, where $\rho$ and $p$ are the energy density and pressure, respectively, which are related through the equation of state and the barotropic index $w$ as $p=w\rho$. Decomposing Eq.~(\ref{eq:bianchi}) into ADM form, it can be shown\footnote{See \cite{ONeal-Ault:2020ebv} for a detailed derivation} that
\begin{equation}\label{eq:bianchispace}
    \nabla_\mu(T_s)^\mu_{~j} \propto \partial_j s_{00},
\end{equation}
and by imposing $\partial_js_{00} = 0$ in the present coordinates, we satisfy part of the contracted Bianchi identities, and we need only treat the component $\nabla_\mu(T)^\mu_{~0}$ further. The $s_{\mu\nu}$ and matter parts are
\begin{align}\label{eq:bianchi0}
\nonumber&\nabla_\mu(T_s)^\mu_{~0} = \frac{\ddot{a}}{a}\left(\tfrac{3}{2}\dot{s}_{00}+6s_{00}\frac{\dot{a}}{a}\right)+3s_{00}\frac{\dddot{a}}{a}, \\
&\nabla_\mu(T_M)^\mu_{~0} = -\dot{\rho}-3\frac{\dot{a}}{a}(\rho+p),
\end{align}
which have been derived by plugging in the metric (\ref{eq:flrw}). It should be noted that energy-momentum conservation of this form is an assumption, and should be interpreted as a choice of model.

We find the Friedmann equations in \cite{ONeal-Ault:2020ebv}, where we performed a Legendre transformation $\mathcal{H}=\pi^{ij}\dot{\gamma}_{ij}+\pi_\alpha \dot{\alpha} - \mathcal{L}$ and derived the Friedmann equations directly from the canonical momenta and their derivatives. Bearing in mind that the 3D curvature scalar is $\mathcal{R}=6k/a^2$, the background Friedmann equations read
\begin{align}
    \left(\frac{\dot{a}}{a}\right)^2(1-s_{00}) =& \frac{\rho}{3}-\frac{k}{a^2}-s_{00}\frac{\ddot{a}}{a}+\frac{\dot{a}}{a}\frac{\dot{s}_{00}}{2} \\
    \left[\frac{\ddot{a}}{a}+\frac{1}{2}\left(\frac{\dot{a}}{a}\right)^2\right](1-s_{00}) =& -\frac{p}{2} - \frac{k}{2a^2}+\frac{\dot{a}}{a}\dot{s}_{00}+\tfrac{1}{4}\ddot{s}_{00},
\end{align}
which can be shown to be consistent with Eq.~(\ref{eq:bianchi0}). Note that we only consider the flat ($k=0$) case, but we write out the Friedmann equations for the general case for completeness. Also, we will \emph{not} impose separate conservation of $(T_s)^{\mu\nu}$ and $(T_M)^{\mu\nu}$, and instead consider the conservation of the stress-energy tensor as a whole, i.e.
\begin{equation}\label{eq:bianchitime}
    \nabla_\mu \left[(T_s)^{\mu}_{~0}+(T_M)^{\mu}_{~0}\right]=0,
\end{equation}
which in general leads to a modified continuity equation for matter fields of the form $\dot{\rho}+3Hf(w,s_{00})\rho=0$, where $H=\dot{a}/a$ is the Hubble parameter. In the special case where $\dot{s}_{00}=0$, the evolution of radiation ($w=1/3$) and the cosmological constant ($w=-1$) are modified, whilst the other matter fields (baryonic matter and curvature) are untouched. The adherence to Eqs.~(\ref{eq:bianchispace}) and (\ref{eq:bianchitime}) ensures compatibility with the contracted Bianchi identities.

\section{Perturbations}\label{sec:perts}
We consider tensor perturbations around a flat FLRW geometry described by the line element 
\begin{equation}\label{eq:flrwflat}
    ds^2 = -dt^2 + a(t)(\delta_{ij}+h_{ij}(t,\mathbf{x}))dx^idx^j,
\end{equation}
where $h_{ij}(t,\mathbf{x})$ is a transverse and traceless spatial tensor perturbation $(h^i_{~i} = \partial^ih_{ij}=0)$. Note that this gauge freedom still remains, since the choice of $s_{\mu\nu}$ only leads to a breaking of \emph{timelike} particle diffeomorphisms. From the background line element (\ref{eq:ADMmetric}), we identify the ADM lapse and shift as $\alpha = 1$, $\beta^i = 0$.
We now perturb the Lagrangian (\ref{eq:Lagrred}) to second order in $h_{ij}$ and find the equation of motion of the perturbation as in \cite{wang}; after some simplification, the quadratic action can be written as
\begin{equation}
    \delta^2S = \tfrac{1}{4}\int dtd^3x a\left[a^2(1-s_{00})\dot{h}_{ij}\dot{h}^{ij}-\partial_kh_{ij}\partial^kh^{ij}\right],
\end{equation}
where an overdot denotes differentiation with respect to cosmic time $t$. It is important to note that no time derivatives of $s_{00}$ appear at second order in perturbation, but do show up at the background level\footnote{If one also considers perturbations of $s_{00}$, the first-order Lagrangian is no longer zero, and extra terms appear at second order.} (zeroth order), which was already noted in~\cite{ONeal-Ault:2021uwu}. It is also worth pointing out here that $s_{00}$ needs to be smaller than unity at all times, to avoid the appearance of tachyonic ghost modes. 
The Euler-Lagrange equation for $h_{ij}$ reads
\begin{equation}
    \ddot{h}_{ij}+\left(3H-\frac{\dot{s}_{00}}{1-s_{00}}\right)\dot{h}_{ij}-\frac{1}{a^2(1-s_{00})}\nabla^2h_{ij}=0,
\end{equation}
where $\nabla^2$ is the Laplace operator.
We now introduce the canonical Fourier-space decomposition of the tensor perturbations as
\begin{equation}\label{eq:fourier}
    h_{ij}(t,\mathbf{x}) = \sum_{\lambda}\int \frac{d^3k}{(2\pi)^{3/2}}\epsilon_{ij}^\lambda(\mathbf{k})h_\mathbf{k}(t)e^{i \mathbf{k}\cdot\mathbf{x}},
\end{equation}
where $\epsilon_{ij}^\lambda$ are the symmetric polarisation tensors obeying $\epsilon^{ij}_\lambda \epsilon_{ij}^{\lambda^\prime} = 2\delta^{~\lambda^\prime}_{\lambda}\delta^{(3)}(\mathbf{k}-\mathbf{k}^\prime)$, and $\lambda=\{+,\times\}$ represents the two propagating polarisation modes. In this way, the equation of motion for the Fourier modes becomes
\begin{equation}\label{eq:eoms}
    \boxed{\ddot{h}_\mathbf{k}+\left(3H-\frac{\dot{s}_{00}}{1-s_{00}}\right)\dot{h}_\mathbf{k}+\frac{k^2}{a^2(1-s_{00})}h_\mathbf{k} = 0},
\end{equation}
which flows smoothly to GR in the limit $s_{00}\rightarrow 0$. Before proceeding, we check that $h_\mathbf{k}$ is a constant outside the horizon ($k \ll aH$); here, we can neglect the third term of Eq.~(\ref{eq:eoms}) (under the important provision that $s_{00}\ll 1$ at all times, as mentioned above) and form
\begin{equation}
    \frac{\ddot{h}_\mathbf{k}}{\dot{h}_\mathbf{k}} \approx \frac{\dot{s}_{00}}{1-s_{00}}-3H,
\end{equation}
which has the general solution
\begin{equation}
    h_\mathbf{k}(t) = \xi_1 + \xi_2\int_1^t\frac{dt^\prime}{(1-s_{00}(t^\prime))a(t^\prime)^3}. 
\end{equation}
The integral part is a decaying mode, which can be neglected (as long as $k \ll aH$ holds for a sufficiently long time), and we can use as as our initial condition $h_\mathbf{k} \rightarrow \text{const.}$ on superhorizon scales. 

For phenomenological purposes, we also refer to a general formula for gravitational-wave propagation (with zero anisotropic stress) from an effective field theory (using again cosmic time $t$) as \cite{Pettorino:2014bka,Nishizawa:2017nef}
\begin{equation}
    \ddot{h}_\mathbf{k}+(3+\mathcal{A})H \dot{h}_\mathbf{k} +\left( c_T^2\frac{k^2}{a^2}+m_g\right)h_\mathbf{k} = 0,
\end{equation}
where $c_T$ is the propagation speed of the tensor modes, $\mathcal{A}=\mathcal{H}^{-1}(d \ln{M_{\rm Pl}^2}/dt)$ is the Planck mass run rate, and $m_g$ is the graviton mass. In what follows, we will be able to relate $\mathcal{A}$ and $m_g$ with $s_{00}$ and its derivatives; without simplifying further, we can immediately read off
\begin{equation}
\begin{aligned}
    c_T^2 = \frac{1}{1-s_{00}}, \quad m_g = 0\\
    \frac{d\ln{M_{\rm Pl}^2}}{dt} = -\frac{\dot{s}_{00}}{1-s_{00}},
\end{aligned}
\end{equation}
both of which show the correct GR limit for $s_{00}\rightarrow 0$. The speed of the tensor modes is not constant, despite a non-zero graviton mass, which suggests some kind of birefringent or spacetime-foam behaviour, accompanied by a modified phase velocity due to the non-zero spacetime-symmetry breaking.\footnote{It should be noted that for constant $s_{00}$, $c_T^2$ can be expanded around zero to resemble the ``EFT-inspired ansatz'' for $c_T^2$ and the ``polynomial ansatz'' for $c_T$ (i.e. no square) used in \cite{Baker:2022rhh}; however, care should be taken to avoid inconsistencies in perturbation order.}. Currently, the best constraint on the ``speed of gravity'' comes from the LIGO/Virgo event GW170817 with associated EM-counterpart GRB170817A, which yielded the constraint $-3\cdot 10^{-15} < c_T -1 < +7\cdot 10^{-16}$. From this constraint we immediately get the following bound $$-6\cdot 10^{-15}<s_{00}<1.4\cdot 10^{-15},$$
    which is similar to other SME bounds from gravitational waves~\cite{Kostelecky:2008ts}. The phenomenology of a running Planck mass has been considered in \cite{Robbers:2007ca,Lagos:2019kds}, and also in scalar-tensor type models such as Einstein-Aether, where it was found that the amplitude of the matter power spectrum is reduced, as compared to the CMB power spectrum; moreover, since the Planck mass influence the amplitude of gravitational waves, the cosmological distance measures will be affected, as measured through gravitational-wave detections \cite{Lagos:2019kds}. Since a varying Planck mass implies a varying Newton's constant, this will affect the Equivalence Principle, from which there are stringent constraints on spacetime-symmetry breaking \cite{Kostelecky:2008in}, most notably from the MICROSCOPE mission which has produced constraints on symmetry-breaking coefficients on the order of $10^{-14}$ \cite{Pihan-LeBars:2019qsd}. Other bounds on spacetime-symmetry breaking from the speed of gravity exists, for example in the context of linearised gravity \cite{Liu:2020slm}. There are also bounds on gravitational birefringence and anisotropic gravitational-dispersion from LIGO-Virgo-Kagra data (GWTC-3 catalogue) on the order of $3.10 \cdot 10^{-15}$ and $10^{-13}$, respectively\cite{Haegel:2022ymk}. 

We now proceed in solving Eq.~(\ref{eq:eoms}) by setting $s_{00}$ to be a constant with respect to \emph{coordinate time} $t$, after which we derive the consequences using conformal time defined through $dt=a d\eta$.\footnote{It can be checked that the superhorizon initial condition holds also in these coordinates.}. Where necessary, we will specify the functional dependence of $s_{00}$, i.e. $s_{00}(t)$.

Restricting $s_{00}$ to be a constant with respect to cosmic time affects the dynamics at both the background and perturbative level; the Friedmann equations can now be written as
\begin{equation}\label{eq:friedt}
    \frac{H^2}{H_0^2} = \Omega_m^0 a^{-3} + \Omega_r^0 a^{-4 x_r} + \Omega_\Lambda^0 a^{-x_\Lambda}+\Omega_k^0 a^{-2},
\end{equation}
where\footnote{A summary and plot of all quantities depending on $s_{00}$ can be found in Appendix~\ref{app:s00dep}.} $x_r = (1-\tfrac{3}{4}s_{00})/(1-\tfrac{1}{2}s_{00})$ and $x_\Lambda=-3s_{00}/(1-\tfrac{5}{2}s_{00})$ arise from the modified continuity equation $\dot{\rho}+3\frac{\dot{a}}{a}f(w,s_{00})\rho=0$, where
\begin{equation}
    f(w,s_{00})=\frac{2(1+w-s_{00})}{2+s_{00}(3w-2)}.
\end{equation}
In these equations, common factors of $s_{00}$ have been absorbed into the definition of the $\Omega$'s, and no rescaling of the time coordinate or the scale factor can fully eliminate its effects. The full set of equations can be seen in Appendix~\ref{app:backevol}.
It is worth pointing out that since we do not have a pure cosmological constant due to the modified evolution, this sector behaves like dynamical dark energy which evolves very slowly according to $\rho_\Lambda = \rho_\Lambda^0 a^{-x_\Lambda}$.

The equation of motion for the Fourier modes simplifies to
\begin{equation}\label{eq:sdot0t}
    \ddot{h}_\mathbf{k}+3H\dot{h}_\mathbf{k}+\frac{k^2}{a^2(1-s_{00})}h_\mathbf{k} = 0,
\end{equation}
where the friction term generated by $\dot{s}_{00}$ now is absent, and the only symmetry-breaking modification appears in the background evolution $H$, as well as in the scaling of the comoving wavenumber. 

Now that we have established the consequences of $\dot{s}_{00}=0$ on the equations of motion, we move to the conformal time coordinate, after which Eq.~(\ref{eq:sdot0t}) reads
\begin{equation}\label{eq:eometa}
    h_\mathbf{k}^{\prime\prime}+2\mathcal{H}h_\mathbf{k}^\prime+\frac{k^2}{1-s_{00}}h_\mathbf{k} = 0.
\end{equation}
In order to facilitate solving this equation, we note that since cosmological evolution is modified at the background level, the standard expressions for $\mathcal{H}$ in de-Sitter and radiation-domination eras are modified. From Eq.~(\ref{eq:friedt}) we find
\begin{equation}\label{eq:confH}
\mathcal{H}_{\rm dS} = -\frac{2}{(2-x_\Lambda)\eta}, \quad \mathcal{H}_{\rm RD}=\frac{1}{(2x_r-1)\eta}.
\end{equation}
In matter domination (MD), the background evolution remains the same, $\mathcal{H}_{\rm MD}=1/2\eta$. 

We now proceed to the power spectrum of the primordial gravitational waves produced during the modified de-Sitter phase; starting from the mode function of tensor perturbations, we write down the quantum operator by introducing creation and annihilation operators as
\begin{equation}
    \hat{\tilde{h}}_\mathbf{k}(\eta) = v_\mathbf{k}(\eta)\hat{a}_\mathbf{k} + v^*_\mathbf{k}(\eta)\hat{a}_\mathbf{k}^\dagger,
\end{equation}
where $\tilde{h}_\mathbf{k}=ah_\mathbf{k}$, and the coefficients $v_\mathbf{k}(\eta)$ satisfy the associated Mukhanov-Sasaki equation, which reads
\begin{equation}\label{eq:MS}
    v_\mathbf{k}^{\prime\prime}+\left(\frac{k}{1-s_{00}}-\frac{2}{\eta^2}\gamma\right)v_\mathbf{k} = 0,
\end{equation}
where we have introduced $\gamma = (4-x_\Lambda)/(2-x_\lambda)^2$ for convenience. The general solutions to this equation are shown in Appendix~\ref{app:gensol}. Note that we could have chosen to rescale $k$ as $\bar{k}^2=k^2/(1-s_{00})$; however, since $s_{00}$ appears both at the background and perturbation level, there will be a second modification appearing when evaluating $a^{\prime\prime}/a$. These two (constant, but undetermined) scalings are coupled though $s_{00}$, and rescaling both terms independently is therefore not possible.
The two-point correlation function reads \cite{Boyle:2005se}
\begin{equation}
    \langle0|\hat{h}_\mathbf{k}^\dagger \hat{h}_\mathbf{k}|0\rangle = \frac{16\pi G}{a^2}\int_0^\infty \frac{dk}{k}\frac{k^3}{2\pi^2}|v_\mathbf{k}|^2,
\end{equation}
from which the power spectrum can be defined\footnote{Several slightly different definitions of $\Delta_h^2$ exist in the literature; here, we adhere to that of \cite{Boyle:2005se}, which also agrees with the \emph{Planck} definition.} as the logarithmic derivative of the tensor modes
\begin{equation}
    \Delta_h^2(\eta,k) \equiv \frac{d\langle0|\hat{h}_\mathbf{k}^\dagger \hat{h}_\mathbf{k}|0\rangle}{d\ln{k}} = \frac{16 \pi G}{a^2} \frac{k^3}{2\pi^2}|v_\mathbf{k}|^2.
\end{equation}
After some simplification, we arrive at the final expression for the power spectrum, which reads
\begin{equation}\label{eq:Powspec}
    \boxed{\Delta_h^2(\eta,k) = \frac{16\pi G}{2\pi^2}H^2(2-x_\Lambda)^2\gamma^2\left|A_\mathbf{k}(\eta)\right|^2},
\end{equation}
where $H=\dot{a}/a$ is the Hubble parameter, which we have taken as constant when deriving this result, and $A_\mathbf{k}(\eta)$ reads
\begin{equation}\label{eq:A}
    A_\mathbf{k}(\eta) = e^{\tfrac{2ik\eta}{1-s_{00}}}\left(1+e^{i\pi\sqrt{1+8\gamma}}\right)-ie^{\tfrac{i\pi}{2}\sqrt{1+8\gamma}},
\end{equation}
which reduces to the  well-known solution based on the Bunch-Davies vacuum in the GR limit $s_{00}\rightarrow 0$, $\gamma \rightarrow 1$, where the Wronskian normalisation condition holds trivially.\footnote{It can also easily be checked that the solution for $v_\mathbf{k}(\eta)$ has the correct GR limit.}
\begin{figure*}[ht]
 \includegraphics[width=0.8\textwidth]{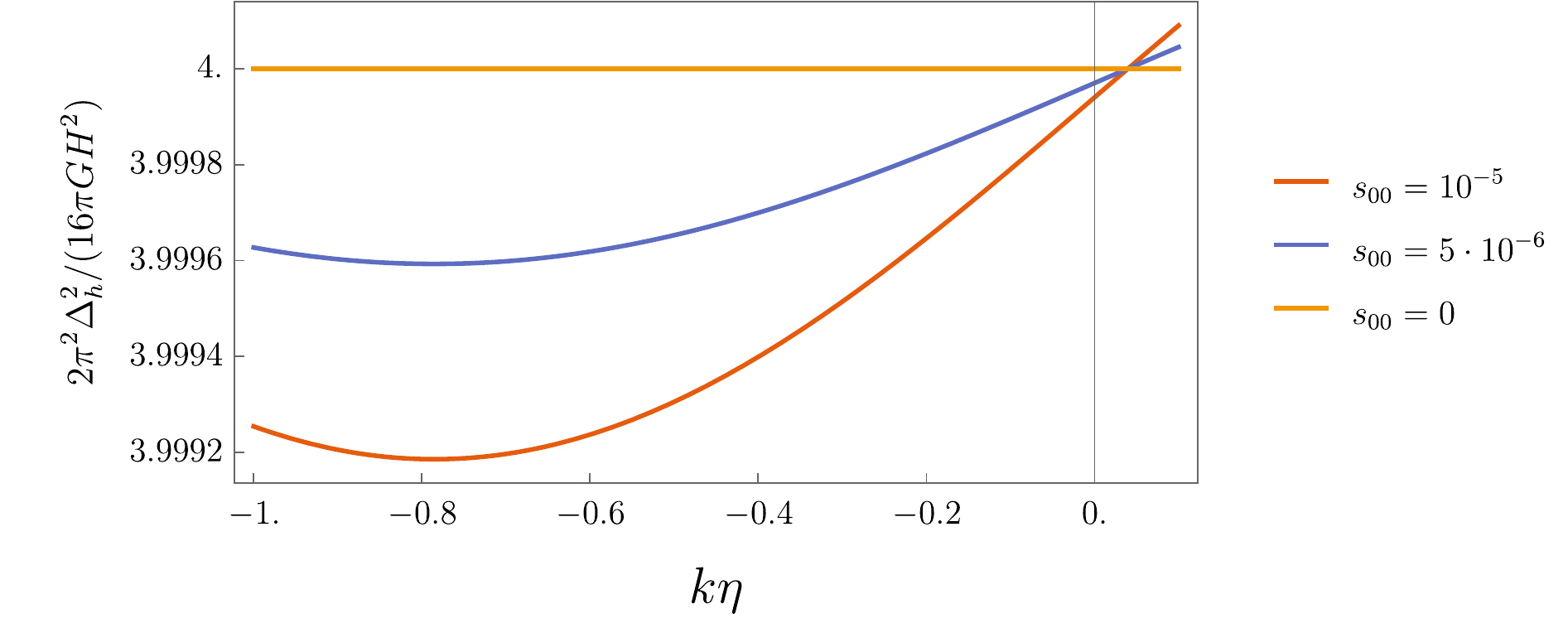}
 \caption{The scale non-invariance in the power spectrum as induced by a non-zero $s_{00}$.}
 \label{fig:scaledep}
\end{figure*}
Therefore, the existence of a non-zero $s_{00}$ \emph{breaks scale invariance} of the power spectrum for modes entering the horizon before the end of inflation, i.e. for $\eta <0$, and the power spectrum takes on a slowly oscillatory behaviour, as can be seen in Figure~\ref{fig:scaledep}. 
\begin{figure}
  \centering
\begin{subfigure}{0.48\textwidth} 
 \includegraphics[width=\textwidth]{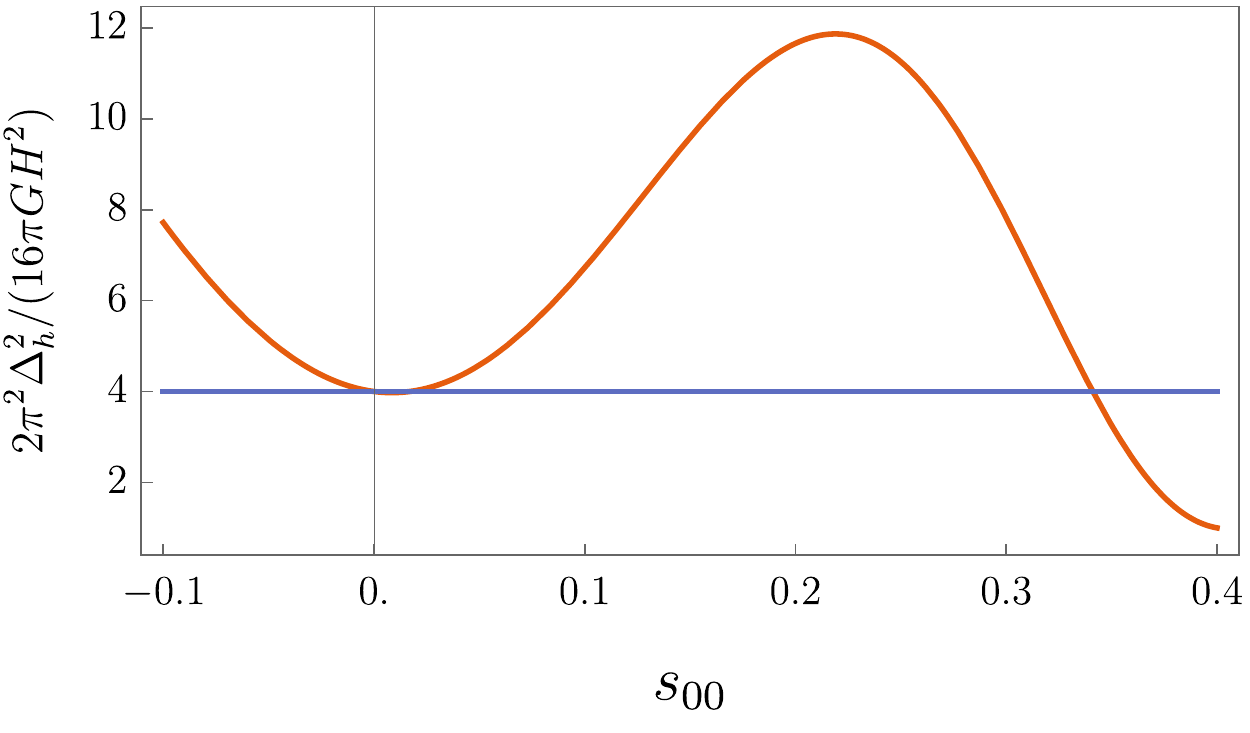}
 \caption{The primordial power spectrum showing that three values of $s_{00}$ give the same amplitude as in GR.}
 \label{fig:Deltah2s00}
\end{subfigure}
\begin{subfigure}{0.48\textwidth}  
 \includegraphics[width=\textwidth]{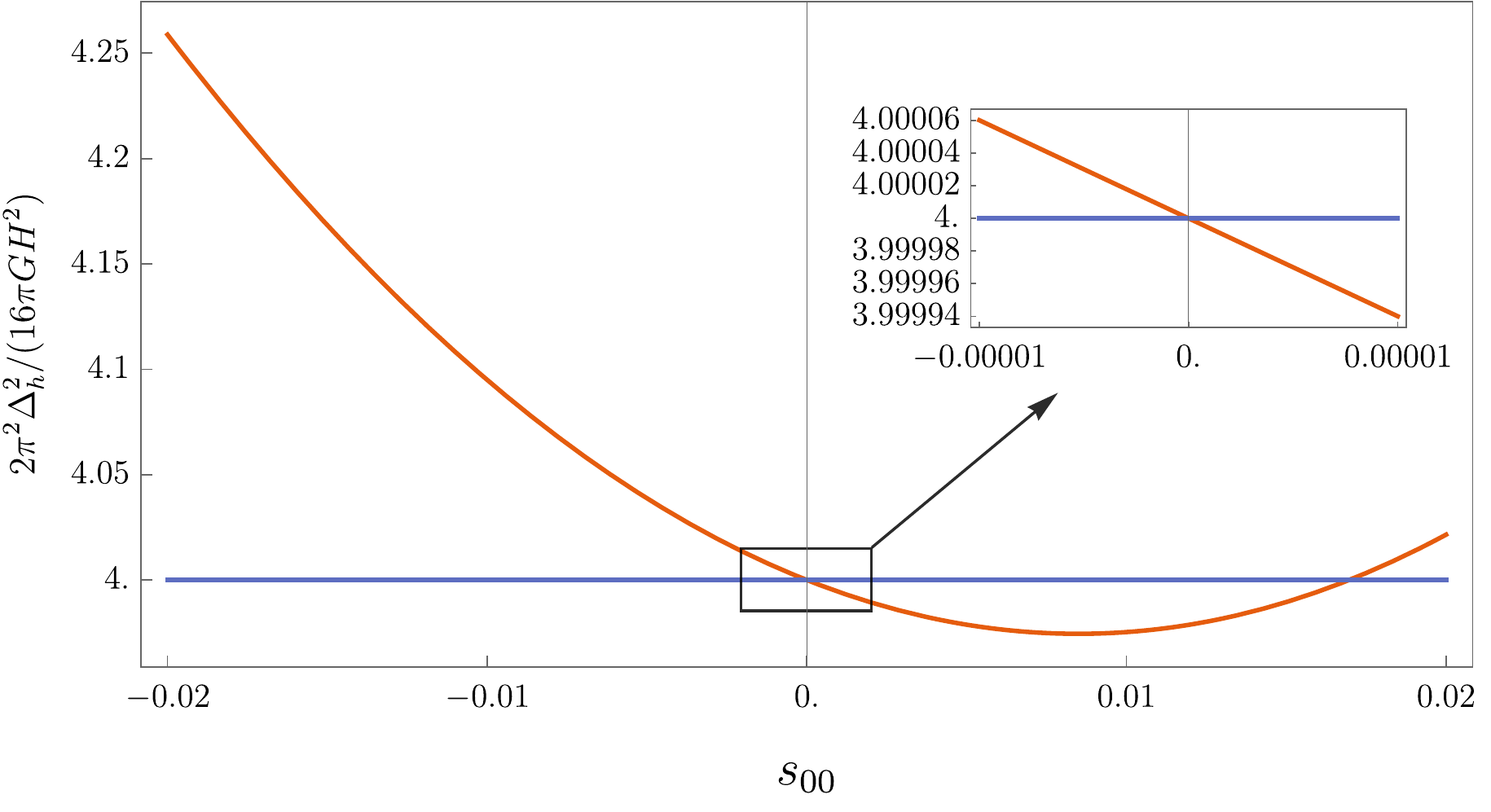}
 \caption{The primordial power spectrum zoomed in around $s_{00}=0$.}
 \label{fig:Deltah2s00zoom}
\end{subfigure}
\caption{The behaviour of the primordial power spectrum defined at $k\eta=0$ as a function of $s_{00}$.}
\label{fig:PS}
\end{figure}
For the \emph{primordial spectrum}, canonically defined exactly at the end of inflation ($\eta=0$), we reobtain the expected scale invariance, but the amplitude is modified through $s_{00}$. Figure~\ref{fig:PS} shows the behaviour of the primordial power spectrum as a function of $s_{00}$.
As can be seen in Figure~\ref{fig:PS}, the value of $s_{00}$ strongly affects the amplitude of the primordial tensor power spectrum, but only at very large, likely unphysical values. From the bound we obtained in in Section III, $|s_{00}| \sim 10^{-15}$, the deviation from the GR case is miniscule; even with a very optimistic estimate of $|s_{00}| \sim 10^{-4}$, the deviation is on the order of $10^{-4} - 10^{-5}$, as can be seen in the inset in Figure~\ref{fig:Deltah2s00zoom}.
\subsubsection{Radiation domination}
For completeness, we also briefly examine the behaviour of the Fourier modes for $\eta>0$, i.e. during radiation domination. In this case, the conformal Hubble parameter is $\mathcal{H}_{\rm RD}$ as given in Eq.~(\ref{eq:confH}), and we solve Eq.~(\ref{eq:eometa}) in the same manner as for de Sitter in the previous section (the general solution is shown in Appendix~\ref{app:gensol}).
\begin{figure}[h]
    \includegraphics[width=0.48\textwidth]{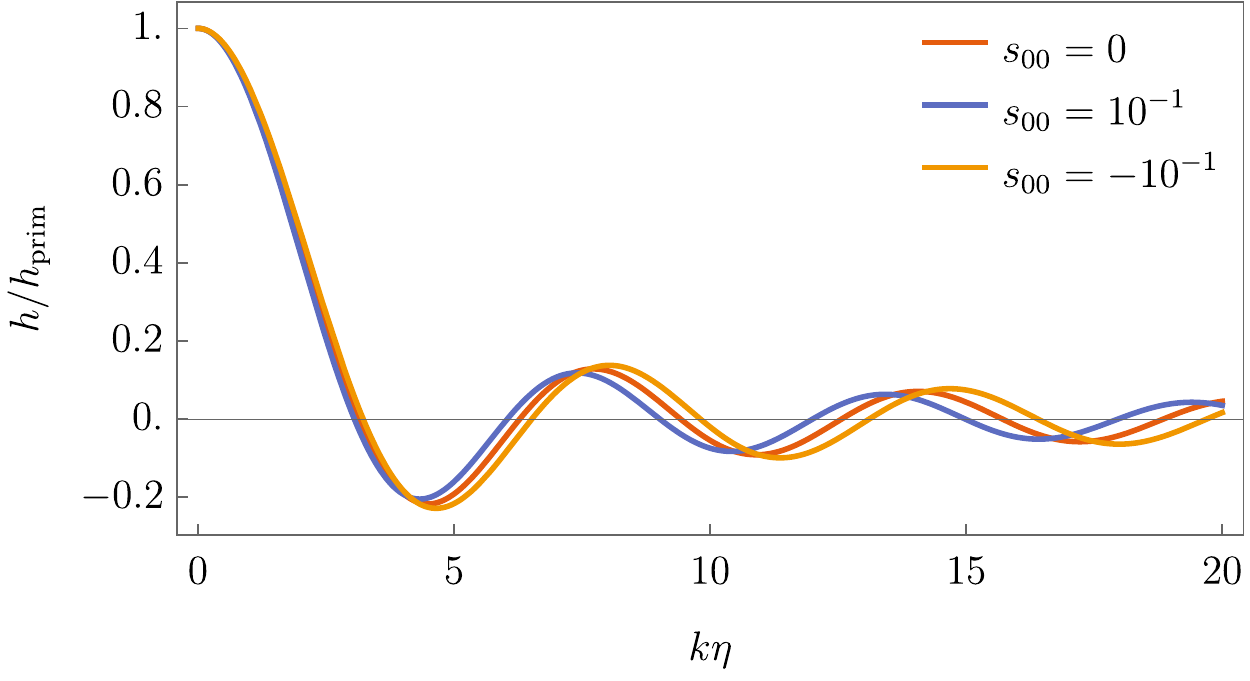}
    \caption{Behaviour of the classical mode function $h_\mathbf{k}$ during radiation domination.}
    \label{fig:RD}
\end{figure}
Figure~\ref{fig:RD} shows the behaviour of the normalised Fourier modes $h_\mathbf{k}$ as a function of $k\eta$: for GR ($s_{00}\rightarrow 0$) we obtain the same solution as in Figure 7 of \cite{Watanabe:2006qe}; turning on $s_{00}$ introduces a \emph{phase shift} in the amplitude (note that we have used very large values of $s_{00}$ in Figure~\ref{fig:RD} in order to highlight this effect).

\section{Discussion}\label{sec:disc}
In this paper we have considered tensor perturbations around flat FLRW spacetime in the presence of a simple case of explicitly spacetime-symmetry breaking terms. By examining solutions to the equations of motion for the Fourier modes and taking into account the fact that spacetime-symmetry breaking alters the continuity equation, we have shown that this produces modified evolution of the modes in both the de Sitter and radiation-domination epochs. Using the current constraints on the speed of gravity, we have found a constraint on the coefficient $s_{00}$ on the order of $10^{-15}$, which is compatible with other constraints from gravitational waves, although very few constraints exist in the exact regime. After quantising the perturbations, we obtain solutions to the Mukhanov-Sasaki equation in the form of a modified Bunch-Davies vacuum, and calculate the primordial power spectrum which exhibits a modified amplitude. By working in the extact regime, we avoid issues with perturbation order in the case of several small quantities; instead perturbation order and order of symmetry breaking can be counted separately, in a similar way as in \cite{Kostelecky:2010ze}. The considerations in this paper constitute some of the first results in the context of early-Universe physics within this framework.

In the future, generalisations beyond the case studied in this paper may be considered for a more complete phenomenological picture which includes quantisation of the tensor perturbations in the spontaneous-breaking case. It is likely that this will provide constraints on spacetime-symmetry breaking from the anisotropies in the CMB, as well as observations with next-generation gravitational wave detectors such as LISA.

\begin{acknowledgments}
We thank Quentin G. Bailey for helpful discussions and feedback, and the anonymous referee who helped improve this manuscript. NAN acknowledges support from United Kingdom Research and Innovation (UKRI) and the Basic Science Research Program through the National Research
Foundation of Korea (NRF) funded by the Ministry of Education, Science and Technology (2020R1A2C1010372, 2020R1A6A1A03047877).
\end{acknowledgments}

\section{Appendices}
\appendix

\section{General solutions}\label{app:gensol}
In the modified de Sitter phase, with Hubble parameter $\mathcal{H}_{\rm dS}$, Eq.~(\ref{eq:MS}) has the following solution
\begin{equation}
\begin{aligned}
    v_\mathbf{k} = &\xi_1 j(\tfrac{1}{2}(\sqrt{1+8\gamma}-1), k\eta) \\+ &\xi_2 y(\tfrac{1}{2}(\sqrt{1+8\gamma}-1), k\eta),
    \end{aligned}
\end{equation}
where $\xi_1$ and $\xi_2$ are constants, and $j$, $y$ are the spherical Bessel functions of the first and second kind, respectively. 

In radiation domination, using instead $\mathcal{H}_{\rm RD}$ and solving for the Fourier modes in Eq.~(\ref{eq:eometa}), we arrive at the following general solution
\begin{equation}
\begin{aligned}
   h_\mathbf{k} = (k\eta)^{1/(2(s_{00}-1))}\Big[&\zeta_1 J\left(\tfrac{1}{2(s_{00}-1)},-\tfrac{k\eta}{\sqrt{1-s_{00}}}\right) \\+&\zeta_2 Y\left(\tfrac{1}{2(s_{00}-1)},-\tfrac{k\eta}{\sqrt{1-s_{00}}}\right)\Big],
\end{aligned}
\end{equation}
where $\zeta_1$ and $\zeta_2$ are constants, and $J$, $Y$ are Bessel functions of the first and second kind, respectively; these are related to the spherical Bessel functions $j$ and $y$ \cite{howlett_1966}.

\section{Background evolution equations}\label{app:backevol}
The full set of equations, when imposing $\nabla_\mu (T_s)^{\mu}_\nu=0$, are the first and second Friedmann equations and the continuity equation. From \cite{ONeal-Ault:2020ebv} we have for a flat universe and constant $s_{00}$)
\begin{equation}
    H^2 = \frac{\rho}{3(1-\tfrac{3}{2}s_{00})}+\frac{p s_{00}}{(2-3s_{00})(1-s_{00})}
\end{equation}
\begin{equation}
    \frac{\ddot{a}}{a}=-\frac{\rho+3p}{6(1-\tfrac{3}{2}s_{00})},
\end{equation}
where $\rho$ and $p$ is the energy density and pressure, respectively, and a non-standard pressure term appears in the first Friedmann equation.  We can rewrite these equations as
\begin{equation}
    \frac{H^2}{H_0^2}=\Omega_m^0 a^{-3}+\Omega_r^0 a^{-4x_r}+\Omega_\Lambda^0 a^{-x_\Lambda}
\end{equation}
and
\begin{align}
   \nonumber \frac{\ddot{a}}{a H_0^2} =-\frac{1}{2}\Omega_m^0 a^{-3} &- \Omega_r^0 \frac{2(1-s_{00})}{2-s_{00}}a^{-4x_r} \\ &+\Omega_\Lambda^0 \frac{2(1-s_{00})}{2-5s_{00}}a^{-x_\Lambda},
\end{align}
where the density parameters are defined as
\begin{equation}
    \Omega_X = \frac{\rho}{3H^2(1-\tfrac{3}{2}s_{00})}\frac{2+(3w-2)s_{00}}{2(1-s_{00})}
\end{equation}
and the continuity equation is modified as
\begin{equation}
    \dot{\rho}+3\frac{\dot{a}}{a}f(w,s_{00})\rho=0, \quad f(w,s_{00})=\frac{2(1+w-s_{00})}{2+s_{00}(3w-2)}. 
\end{equation}

\section{$s_{00}$-dependent quantities}\label{app:s00dep}
\begin{figure*}
 \includegraphics[width=0.8\textwidth]{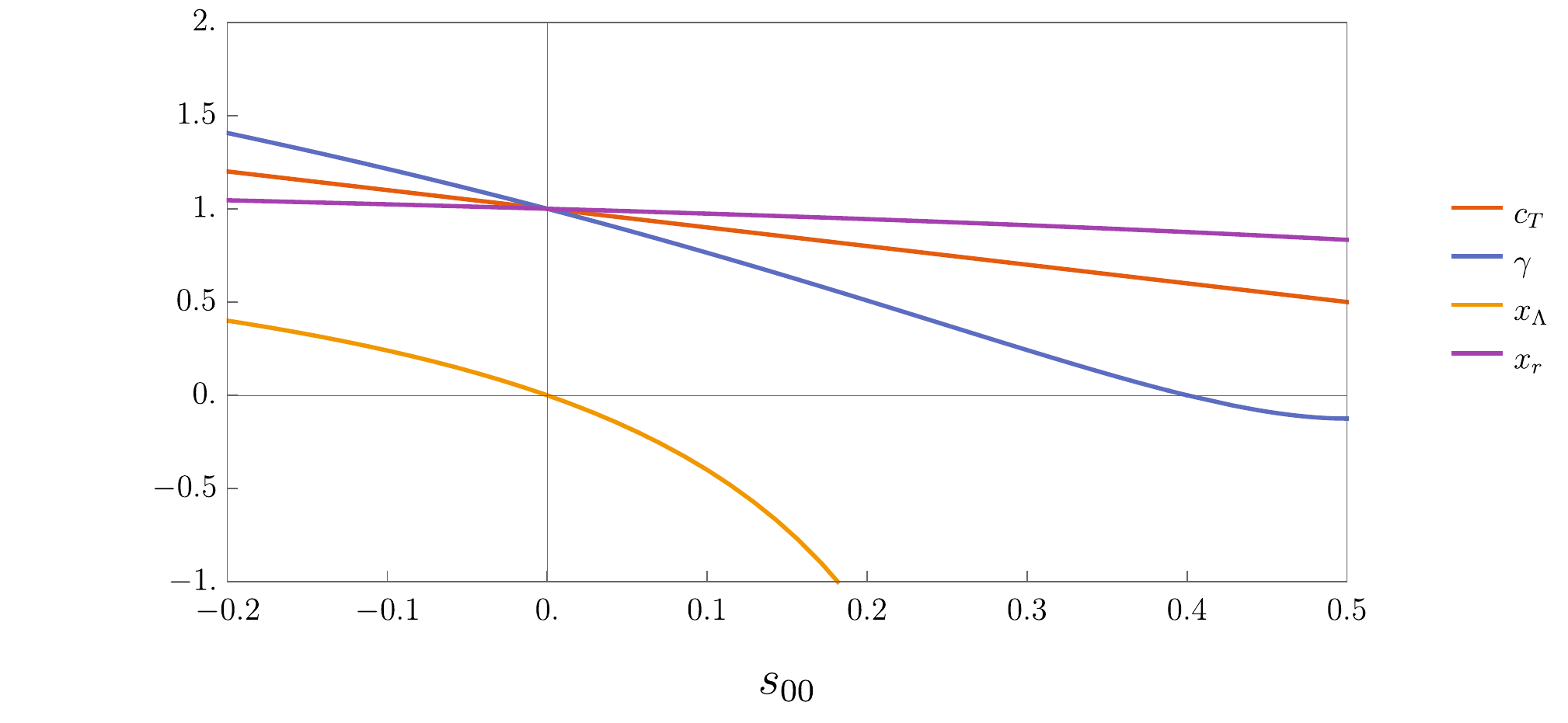}
 \caption{Behaviour of the auxiliary functions.}
 \label{fig:funccomp}
\end{figure*}

Several auxiliary quantities which depend on $s_{00}$ are defined throughout the paper: to recap, they are
\begin{itemize}
    \item $c_T = 1/(1-s_{00})$, the speed of tensor modes.
    \item $x_\Lambda=-3s_{00}/(1-5s_{00}/2)$, the modification to the evolution of the cosmological constant.
    \item $x_r=(1-3s_{00}/4)/(1-s_{00}/2)$, the modification to the evolution of the radiation density.
    \item $\gamma=(4-x_\Lambda)/(2-x_\Lambda)^2$, the modification in the Mukhanov-Sasaki equation (\ref{eq:MS}).
\end{itemize}
The behaviour of these functions with respect to $s_{00}$ can be seen in Figure~\ref{fig:funccomp}

\bibliography{apssamp}

\providecommand{\noopsort}[1]{}\providecommand{\singleletter}[1]{#1}%
\begin{thebibliography}{64}%
\makeatletter
\providecommand \@ifxundefined [1]{%
 \@ifx{#1\undefined}
}%
\providecommand \@ifnum [1]{%
 \ifnum #1\expandafter \@firstoftwo
 \else \expandafter \@secondoftwo
 \fi
}%
\providecommand \@ifx [1]{%
 \ifx #1\expandafter \@firstoftwo
 \else \expandafter \@secondoftwo
 \fi
}%
\providecommand \natexlab [1]{#1}%
\providecommand \enquote  [1]{``#1''}%
\providecommand \bibnamefont  [1]{#1}%
\providecommand \bibfnamefont [1]{#1}%
\providecommand \citenamefont [1]{#1}%
\providecommand \href@noop [0]{\@secondoftwo}%
\providecommand \href [0]{\begingroup \@sanitize@url \@href}%
\providecommand \@href[1]{\@@startlink{#1}\@@href}%
\providecommand \@@href[1]{\endgroup#1\@@endlink}%
\providecommand \@sanitize@url [0]{\catcode `\\12\catcode `\$12\catcode
  `\&12\catcode `\#12\catcode `\^12\catcode `\_12\catcode `\%12\relax}%
\providecommand \@@startlink[1]{}%
\providecommand \@@endlink[0]{}%
\providecommand \url  [0]{\begingroup\@sanitize@url \@url }%
\providecommand \@url [1]{\endgroup\@href {#1}{\urlprefix }}%
\providecommand \urlprefix  [0]{URL }%
\providecommand \Eprint [0]{\href }%
\providecommand \doibase [0]{https://doi.org/}%
\providecommand \selectlanguage [0]{\@gobble}%
\providecommand \bibinfo  [0]{\@secondoftwo}%
\providecommand \bibfield  [0]{\@secondoftwo}%
\providecommand \translation [1]{[#1]}%
\providecommand \BibitemOpen [0]{}%
\providecommand \bibitemStop [0]{}%
\providecommand \bibitemNoStop [0]{.\EOS\space}%
\providecommand \EOS [0]{\spacefactor3000\relax}%
\providecommand \BibitemShut  [1]{\csname bibitem#1\endcsname}%
\let\auto@bib@innerbib\@empty
\bibitem [{\citenamefont {Grishchuk}(1974)}]{Grishchuk:1974ny}%
  \BibitemOpen
  \bibfield  {author} {\bibinfo {author} {\bibfnamefont {L.~P.}\ \bibnamefont
  {Grishchuk}},\ }\href@noop {} {\bibfield  {journal} {\bibinfo  {journal} {Zh.
  Eksp. Teor. Fiz.}\ }\textbf {\bibinfo {volume} {67}},\ \bibinfo {pages} {825}
  (\bibinfo {year} {1974})}\BibitemShut {NoStop}%
\bibitem [{\citenamefont {Starobinsky}(1979)}]{Starobinsky:1979ty}%
  \BibitemOpen
  \bibfield  {author} {\bibinfo {author} {\bibfnamefont {A.~A.}\ \bibnamefont
  {Starobinsky}},\ }\href@noop {} {\bibfield  {journal} {\bibinfo  {journal}
  {JETP Lett.}\ }\textbf {\bibinfo {volume} {30}},\ \bibinfo {pages} {682}
  (\bibinfo {year} {1979})}\BibitemShut {NoStop}%
\bibitem [{\citenamefont {Ma\l{}kiewicz}\ and\ \citenamefont
  {Miroszewski}(2021)}]{Malkiewicz:2020fvy}%
  \BibitemOpen
  \bibfield  {author} {\bibinfo {author} {\bibfnamefont {P.}~\bibnamefont
  {Ma\l{}kiewicz}}\ and\ \bibinfo {author} {\bibfnamefont {A.}~\bibnamefont
  {Miroszewski}},\ }\href {https://doi.org/10.1103/PhysRevD.103.083529}
  {\bibfield  {journal} {\bibinfo  {journal} {Phys. Rev. D}\ }\textbf {\bibinfo
  {volume} {103}},\ \bibinfo {pages} {083529} (\bibinfo {year} {2021})},\
  \Eprint {https://arxiv.org/abs/2011.03487} {arXiv:2011.03487 [gr-qc]}
  \BibitemShut {NoStop}%
\bibitem [{\citenamefont {Kosowsky}\ \emph {et~al.}(1992)\citenamefont
  {Kosowsky}, \citenamefont {Turner},\ and\ \citenamefont
  {Watkins}}]{Kosowsky:1991ua}%
  \BibitemOpen
  \bibfield  {author} {\bibinfo {author} {\bibfnamefont {A.}~\bibnamefont
  {Kosowsky}}, \bibinfo {author} {\bibfnamefont {M.~S.}\ \bibnamefont
  {Turner}},\ and\ \bibinfo {author} {\bibfnamefont {R.}~\bibnamefont
  {Watkins}},\ }\href {https://doi.org/10.1103/PhysRevD.45.4514} {\bibfield
  {journal} {\bibinfo  {journal} {Phys. Rev. D}\ }\textbf {\bibinfo {volume}
  {45}},\ \bibinfo {pages} {4514} (\bibinfo {year} {1992})}\BibitemShut
  {NoStop}%
\bibitem [{\citenamefont {Kamionkowski}\ \emph {et~al.}(1994)\citenamefont
  {Kamionkowski}, \citenamefont {Kosowsky},\ and\ \citenamefont
  {Turner}}]{Kamionkowski:1993fg}%
  \BibitemOpen
  \bibfield  {author} {\bibinfo {author} {\bibfnamefont {M.}~\bibnamefont
  {Kamionkowski}}, \bibinfo {author} {\bibfnamefont {A.}~\bibnamefont
  {Kosowsky}},\ and\ \bibinfo {author} {\bibfnamefont {M.~S.}\ \bibnamefont
  {Turner}},\ }\href {https://doi.org/10.1103/PhysRevD.49.2837} {\bibfield
  {journal} {\bibinfo  {journal} {Phys. Rev. D}\ }\textbf {\bibinfo {volume}
  {49}},\ \bibinfo {pages} {2837} (\bibinfo {year} {1994})},\ \Eprint
  {https://arxiv.org/abs/astro-ph/9310044} {arXiv:astro-ph/9310044}
  \BibitemShut {NoStop}%
\bibitem [{\citenamefont {Easther}\ and\ \citenamefont
  {Lim}(2006)}]{Easther:2006gt}%
  \BibitemOpen
  \bibfield  {author} {\bibinfo {author} {\bibfnamefont {R.}~\bibnamefont
  {Easther}}\ and\ \bibinfo {author} {\bibfnamefont {E.~A.}\ \bibnamefont
  {Lim}},\ }\href {https://doi.org/10.1088/1475-7516/2006/04/010} {\bibfield
  {journal} {\bibinfo  {journal} {JCAP}\ }\textbf {\bibinfo {volume} {04}},\
  \bibinfo {pages} {010}},\ \Eprint {https://arxiv.org/abs/astro-ph/0601617}
  {arXiv:astro-ph/0601617} \BibitemShut {NoStop}%
\bibitem [{\citenamefont {Khlebnikov}\ and\ \citenamefont
  {Tkachev}(1997)}]{Khlebnikov:1997di}%
  \BibitemOpen
  \bibfield  {author} {\bibinfo {author} {\bibfnamefont {S.~Y.}\ \bibnamefont
  {Khlebnikov}}\ and\ \bibinfo {author} {\bibfnamefont {I.~I.}\ \bibnamefont
  {Tkachev}},\ }\href {https://doi.org/10.1103/PhysRevD.56.653} {\bibfield
  {journal} {\bibinfo  {journal} {Phys. Rev. D}\ }\textbf {\bibinfo {volume}
  {56}},\ \bibinfo {pages} {653} (\bibinfo {year} {1997})},\ \Eprint
  {https://arxiv.org/abs/hep-ph/9701423} {arXiv:hep-ph/9701423} \BibitemShut
  {NoStop}%
\bibitem [{\citenamefont {Vachaspati}\ and\ \citenamefont
  {Vilenkin}(1985)}]{Vachaspati:1984gt}%
  \BibitemOpen
  \bibfield  {author} {\bibinfo {author} {\bibfnamefont {T.}~\bibnamefont
  {Vachaspati}}\ and\ \bibinfo {author} {\bibfnamefont {A.}~\bibnamefont
  {Vilenkin}},\ }\href {https://doi.org/10.1103/PhysRevD.31.3052} {\bibfield
  {journal} {\bibinfo  {journal} {Phys. Rev. D}\ }\textbf {\bibinfo {volume}
  {31}},\ \bibinfo {pages} {3052} (\bibinfo {year} {1985})}\BibitemShut
  {NoStop}%
\bibitem [{\citenamefont {Figueroa}\ \emph {et~al.}(2013)\citenamefont
  {Figueroa}, \citenamefont {Hindmarsh},\ and\ \citenamefont
  {Urrestilla}}]{Figueroa:2012kw}%
  \BibitemOpen
  \bibfield  {author} {\bibinfo {author} {\bibfnamefont {D.~G.}\ \bibnamefont
  {Figueroa}}, \bibinfo {author} {\bibfnamefont {M.}~\bibnamefont
  {Hindmarsh}},\ and\ \bibinfo {author} {\bibfnamefont {J.}~\bibnamefont
  {Urrestilla}},\ }\href {https://doi.org/10.1103/PhysRevLett.110.101302}
  {\bibfield  {journal} {\bibinfo  {journal} {Phys. Rev. Lett.}\ }\textbf
  {\bibinfo {volume} {110}},\ \bibinfo {pages} {101302} (\bibinfo {year}
  {2013})},\ \Eprint {https://arxiv.org/abs/1212.5458} {arXiv:1212.5458
  [astro-ph.CO]} \BibitemShut {NoStop}%
\bibitem [{\citenamefont {Garcia-Bellido}\ \emph {et~al.}(1996)\citenamefont
  {Garcia-Bellido}, \citenamefont {Linde},\ and\ \citenamefont
  {Wands}}]{Garcia-Bellido:1996mdl}%
  \BibitemOpen
  \bibfield  {author} {\bibinfo {author} {\bibfnamefont {J.}~\bibnamefont
  {Garcia-Bellido}}, \bibinfo {author} {\bibfnamefont {A.~D.}\ \bibnamefont
  {Linde}},\ and\ \bibinfo {author} {\bibfnamefont {D.}~\bibnamefont {Wands}},\
  }\href {https://doi.org/10.1103/PhysRevD.54.6040} {\bibfield  {journal}
  {\bibinfo  {journal} {Phys. Rev. D}\ }\textbf {\bibinfo {volume} {54}},\
  \bibinfo {pages} {6040} (\bibinfo {year} {1996})},\ \Eprint
  {https://arxiv.org/abs/astro-ph/9605094} {arXiv:astro-ph/9605094}
  \BibitemShut {NoStop}%
\bibitem [{\citenamefont {Dong}\ \emph {et~al.}(2016)\citenamefont {Dong},
  \citenamefont {Kinney},\ and\ \citenamefont {Stojkovic}}]{Dong:2015yjs}%
  \BibitemOpen
  \bibfield  {author} {\bibinfo {author} {\bibfnamefont {R.}~\bibnamefont
  {Dong}}, \bibinfo {author} {\bibfnamefont {W.~H.}\ \bibnamefont {Kinney}},\
  and\ \bibinfo {author} {\bibfnamefont {D.}~\bibnamefont {Stojkovic}},\ }\href
  {https://doi.org/10.1088/1475-7516/2016/10/034} {\bibfield  {journal}
  {\bibinfo  {journal} {JCAP}\ }\textbf {\bibinfo {volume} {10}},\ \bibinfo
  {pages} {034}},\ \Eprint {https://arxiv.org/abs/1511.05642} {arXiv:1511.05642
  [astro-ph.CO]} \BibitemShut {NoStop}%
\bibitem [{\citenamefont {Crowder}\ and\ \citenamefont
  {Cornish}(2005)}]{PhysRevD.72.083005}%
  \BibitemOpen
  \bibfield  {author} {\bibinfo {author} {\bibfnamefont {J.}~\bibnamefont
  {Crowder}}\ and\ \bibinfo {author} {\bibfnamefont {N.~J.}\ \bibnamefont
  {Cornish}},\ }\href {https://doi.org/10.1103/PhysRevD.72.083005} {\bibfield
  {journal} {\bibinfo  {journal} {Phys. Rev. D}\ }\textbf {\bibinfo {volume}
  {72}},\ \bibinfo {pages} {083005} (\bibinfo {year} {2005})}\BibitemShut
  {NoStop}%
\bibitem [{\citenamefont {Kawamura}\ \emph {et~al.}(2011)\citenamefont
  {Kawamura} \emph {et~al.}}]{Kawamura:2011zz}%
  \BibitemOpen
  \bibfield  {author} {\bibinfo {author} {\bibfnamefont {S.}~\bibnamefont
  {Kawamura}} \emph {et~al.},\ }\href
  {https://doi.org/10.1088/0264-9381/28/9/094011} {\bibfield  {journal}
  {\bibinfo  {journal} {Class. Quant. Grav.}\ }\textbf {\bibinfo {volume}
  {28}},\ \bibinfo {pages} {094011} (\bibinfo {year} {2011})}\BibitemShut
  {NoStop}%
\bibitem [{\citenamefont {Mandic}\ and\ \citenamefont
  {Buonanno}(2006)}]{PhysRevD.73.063008}%
  \BibitemOpen
  \bibfield  {author} {\bibinfo {author} {\bibfnamefont {V.}~\bibnamefont
  {Mandic}}\ and\ \bibinfo {author} {\bibfnamefont {A.}~\bibnamefont
  {Buonanno}},\ }\href {https://doi.org/10.1103/PhysRevD.73.063008} {\bibfield
  {journal} {\bibinfo  {journal} {Phys. Rev. D}\ }\textbf {\bibinfo {volume}
  {73}},\ \bibinfo {pages} {063008} (\bibinfo {year} {2006})}\BibitemShut
  {NoStop}%
\bibitem [{\citenamefont {Bernal}\ and\ \citenamefont
  {Hajkarim}(2019)}]{Bernal:2019lpc}%
  \BibitemOpen
  \bibfield  {author} {\bibinfo {author} {\bibfnamefont {N.}~\bibnamefont
  {Bernal}}\ and\ \bibinfo {author} {\bibfnamefont {F.}~\bibnamefont
  {Hajkarim}},\ }\href {https://doi.org/10.1103/PhysRevD.100.063502} {\bibfield
   {journal} {\bibinfo  {journal} {Phys. Rev. D}\ }\textbf {\bibinfo {volume}
  {100}},\ \bibinfo {pages} {063502} (\bibinfo {year} {2019})},\ \Eprint
  {https://arxiv.org/abs/1905.10410} {arXiv:1905.10410 [astro-ph.CO]}
  \BibitemShut {NoStop}%
\bibitem [{\citenamefont {Fabbri}\ and\ \citenamefont
  {Pollock}(1983)}]{Fabbri:1983us}%
  \BibitemOpen
  \bibfield  {author} {\bibinfo {author} {\bibfnamefont {R.}~\bibnamefont
  {Fabbri}}\ and\ \bibinfo {author} {\bibfnamefont {M.~d.}\ \bibnamefont
  {Pollock}},\ }\href {https://doi.org/10.1016/0370-2693(83)91322-9} {\bibfield
   {journal} {\bibinfo  {journal} {Phys. Lett. B}\ }\textbf {\bibinfo {volume}
  {125}},\ \bibinfo {pages} {445} (\bibinfo {year} {1983})}\BibitemShut
  {NoStop}%
\bibitem [{\citenamefont {White}(1992)}]{White:1992fj}%
  \BibitemOpen
  \bibfield  {author} {\bibinfo {author} {\bibfnamefont {M.~J.}\ \bibnamefont
  {White}},\ }\href {https://doi.org/10.1103/PhysRevD.46.4198} {\bibfield
  {journal} {\bibinfo  {journal} {Phys. Rev. D}\ }\textbf {\bibinfo {volume}
  {46}},\ \bibinfo {pages} {4198} (\bibinfo {year} {1992})},\ \Eprint
  {https://arxiv.org/abs/hep-ph/9207239} {arXiv:hep-ph/9207239} \BibitemShut
  {NoStop}%
\bibitem [{\citenamefont {Sahni}(1990)}]{Sahni:1990tx}%
  \BibitemOpen
  \bibfield  {author} {\bibinfo {author} {\bibfnamefont {V.}~\bibnamefont
  {Sahni}},\ }\href {https://doi.org/10.1103/PhysRevD.42.453} {\bibfield
  {journal} {\bibinfo  {journal} {Phys. Rev. D}\ }\textbf {\bibinfo {volume}
  {42}},\ \bibinfo {pages} {453} (\bibinfo {year} {1990})}\BibitemShut
  {NoStop}%
\bibitem [{\citenamefont {Kosteleck\'y}\ and\ \citenamefont
  {Samuel}(1989)}]{KostelStuart1989}%
  \BibitemOpen
  \bibfield  {author} {\bibinfo {author} {\bibfnamefont {V.~A.}\ \bibnamefont
  {Kosteleck\'y}}\ and\ \bibinfo {author} {\bibfnamefont {S.}~\bibnamefont
  {Samuel}},\ }\href {https://doi.org/10.1103/PhysRevD.39.683} {\bibfield
  {journal} {\bibinfo  {journal} {Phys. Rev. D}\ }\textbf {\bibinfo {volume}
  {39}},\ \bibinfo {pages} {683} (\bibinfo {year} {1989})}\BibitemShut
  {NoStop}%
\bibitem [{\citenamefont {Ellis}\ \emph {et~al.}(2000)\citenamefont {Ellis},
  \citenamefont {Mavromatos},\ and\ \citenamefont {Nanopoulos}}]{Ellis:1999jf}%
  \BibitemOpen
  \bibfield  {author} {\bibinfo {author} {\bibfnamefont {J.~R.}\ \bibnamefont
  {Ellis}}, \bibinfo {author} {\bibfnamefont {N.~E.}\ \bibnamefont
  {Mavromatos}},\ and\ \bibinfo {author} {\bibfnamefont {D.~V.}\ \bibnamefont
  {Nanopoulos}},\ }\href {https://doi.org/10.1103/PhysRevD.61.027503}
  {\bibfield  {journal} {\bibinfo  {journal} {Phys. Rev. D}\ }\textbf {\bibinfo
  {volume} {61}},\ \bibinfo {pages} {027503} (\bibinfo {year} {2000})},\
  \Eprint {https://arxiv.org/abs/gr-qc/9906029} {arXiv:gr-qc/9906029}
  \BibitemShut {NoStop}%
\bibitem [{\citenamefont {Gambini}\ and\ \citenamefont
  {Pullin}(1999)}]{Gambini:1998it}%
  \BibitemOpen
  \bibfield  {author} {\bibinfo {author} {\bibfnamefont {R.}~\bibnamefont
  {Gambini}}\ and\ \bibinfo {author} {\bibfnamefont {J.}~\bibnamefont
  {Pullin}},\ }\href {https://doi.org/10.1103/PhysRevD.59.124021} {\bibfield
  {journal} {\bibinfo  {journal} {Phys. Rev. D}\ }\textbf {\bibinfo {volume}
  {59}},\ \bibinfo {pages} {124021} (\bibinfo {year} {1999})},\ \Eprint
  {https://arxiv.org/abs/gr-qc/9809038} {arXiv:gr-qc/9809038} \BibitemShut
  {NoStop}%
\bibitem [{\citenamefont {Alfaro}\ \emph {et~al.}(2000)\citenamefont {Alfaro},
  \citenamefont {Morales-Tecotl},\ and\ \citenamefont
  {Urrutia}}]{Alfaro:1999wd}%
  \BibitemOpen
  \bibfield  {author} {\bibinfo {author} {\bibfnamefont {J.}~\bibnamefont
  {Alfaro}}, \bibinfo {author} {\bibfnamefont {H.~A.}\ \bibnamefont
  {Morales-Tecotl}},\ and\ \bibinfo {author} {\bibfnamefont {L.~F.}\
  \bibnamefont {Urrutia}},\ }\href
  {https://doi.org/10.1103/PhysRevLett.84.2318} {\bibfield  {journal} {\bibinfo
   {journal} {Phys. Rev. Lett.}\ }\textbf {\bibinfo {volume} {84}},\ \bibinfo
  {pages} {2318} (\bibinfo {year} {2000})},\ \Eprint
  {https://arxiv.org/abs/gr-qc/9909079} {arXiv:gr-qc/9909079} \BibitemShut
  {NoStop}%
\bibitem [{\citenamefont {Alfaro}\ \emph {et~al.}(2002)\citenamefont {Alfaro},
  \citenamefont {Morales-Tecotl},\ and\ \citenamefont
  {Urrutia}}]{Alfaro:2001rb}%
  \BibitemOpen
  \bibfield  {author} {\bibinfo {author} {\bibfnamefont {J.}~\bibnamefont
  {Alfaro}}, \bibinfo {author} {\bibfnamefont {H.~A.}\ \bibnamefont
  {Morales-Tecotl}},\ and\ \bibinfo {author} {\bibfnamefont {L.~F.}\
  \bibnamefont {Urrutia}},\ }\href {https://doi.org/10.1103/PhysRevD.65.103509}
  {\bibfield  {journal} {\bibinfo  {journal} {Phys. Rev. D}\ }\textbf {\bibinfo
  {volume} {65}},\ \bibinfo {pages} {103509} (\bibinfo {year} {2002})},\
  \Eprint {https://arxiv.org/abs/hep-th/0108061} {arXiv:hep-th/0108061}
  \BibitemShut {NoStop}%
\bibitem [{\citenamefont {Kosteleck\'y}(2004)}]{PhysRevD.69.105009}%
  \BibitemOpen
  \bibfield  {author} {\bibinfo {author} {\bibfnamefont {V.~A.}\ \bibnamefont
  {Kosteleck\'y}},\ }\href {https://doi.org/10.1103/PhysRevD.69.105009}
  {\bibfield  {journal} {\bibinfo  {journal} {Phys. Rev. D}\ }\textbf {\bibinfo
  {volume} {69}},\ \bibinfo {pages} {105009} (\bibinfo {year}
  {2004})}\BibitemShut {NoStop}%
\bibitem [{\citenamefont {Colladay}\ and\ \citenamefont
  {Kosteleck\'y}(1998)}]{PhysRevD.58.116002}%
  \BibitemOpen
  \bibfield  {author} {\bibinfo {author} {\bibfnamefont {D.}~\bibnamefont
  {Colladay}}\ and\ \bibinfo {author} {\bibfnamefont {V.~A.}\ \bibnamefont
  {Kosteleck\'y}},\ }\href {https://doi.org/10.1103/PhysRevD.58.116002}
  {\bibfield  {journal} {\bibinfo  {journal} {Phys. Rev. D}\ }\textbf {\bibinfo
  {volume} {58}},\ \bibinfo {pages} {116002} (\bibinfo {year}
  {1998})}\BibitemShut {NoStop}%
\bibitem [{\citenamefont {Colladay}\ and\ \citenamefont
  {Kosteleck\'y}(1997)}]{PhysRevD.55.6760}%
  \BibitemOpen
  \bibfield  {author} {\bibinfo {author} {\bibfnamefont {D.}~\bibnamefont
  {Colladay}}\ and\ \bibinfo {author} {\bibfnamefont {V.~A.}\ \bibnamefont
  {Kosteleck\'y}},\ }\href {https://doi.org/10.1103/PhysRevD.55.6760}
  {\bibfield  {journal} {\bibinfo  {journal} {Phys. Rev. D}\ }\textbf {\bibinfo
  {volume} {55}},\ \bibinfo {pages} {6760} (\bibinfo {year}
  {1997})}\BibitemShut {NoStop}%
\bibitem [{\citenamefont {Iorio}(2012)}]{Iorio:2012gr}%
  \BibitemOpen
  \bibfield  {author} {\bibinfo {author} {\bibfnamefont {L.}~\bibnamefont
  {Iorio}},\ }\href {https://doi.org/10.1088/0264-9381/29/17/175007} {\bibfield
   {journal} {\bibinfo  {journal} {Class. Quant. Grav.}\ }\textbf {\bibinfo
  {volume} {29}},\ \bibinfo {pages} {175007} (\bibinfo {year} {2012})},\
  \Eprint {https://arxiv.org/abs/1203.1859} {arXiv:1203.1859 [gr-qc]}
  \BibitemShut {NoStop}%
\bibitem [{\citenamefont {HEES}\ \emph {et~al.}()\citenamefont {HEES},
  \citenamefont {LAMINE}, \citenamefont {REYNAUD}, \citenamefont {JAEKEL},
  \citenamefont {PONCIN-LAFITTE}, \citenamefont {LAINEY}, \citenamefont
  {FÜZFA}, \citenamefont {COURTY}, \citenamefont {DEHANT},\ and\ \citenamefont
  {WOLF}}]{doi:10.1142/9789814623995_0440}%
  \BibitemOpen
  \bibfield  {author} {\bibinfo {author} {\bibfnamefont {A.}~\bibnamefont
  {HEES}}, \bibinfo {author} {\bibfnamefont {B.}~\bibnamefont {LAMINE}},
  \bibinfo {author} {\bibfnamefont {S.}~\bibnamefont {REYNAUD}}, \bibinfo
  {author} {\bibfnamefont {M.-T.}\ \bibnamefont {JAEKEL}}, \bibinfo {author}
  {\bibfnamefont {C.~L.}\ \bibnamefont {PONCIN-LAFITTE}}, \bibinfo {author}
  {\bibfnamefont {V.}~\bibnamefont {LAINEY}}, \bibinfo {author} {\bibfnamefont
  {A.}~\bibnamefont {FÜZFA}}, \bibinfo {author} {\bibfnamefont {J.-M.}\
  \bibnamefont {COURTY}}, \bibinfo {author} {\bibfnamefont {V.}~\bibnamefont
  {DEHANT}},\ and\ \bibinfo {author} {\bibfnamefont {P.}~\bibnamefont {WOLF}},\
  }\bibinfo {title} {Simulations of solar system observations in alternative
  theories of gravity},\ in\ \href {https://doi.org/10.1142/9789814623995_0440}
  {\emph {\bibinfo {booktitle} {The Thirteenth Marcel Grossmann Meeting}}},\
  pp.\ \bibinfo {pages} {2357--2359}\BibitemShut {NoStop}%
\bibitem [{\citenamefont {Le~Poncin-Lafitte}\ \emph {et~al.}(2016)\citenamefont
  {Le~Poncin-Lafitte}, \citenamefont {Hees},\ and\ \citenamefont
  {Lambert}}]{PhysRevD.94.125030}%
  \BibitemOpen
  \bibfield  {author} {\bibinfo {author} {\bibfnamefont {C.}~\bibnamefont
  {Le~Poncin-Lafitte}}, \bibinfo {author} {\bibfnamefont {A.}~\bibnamefont
  {Hees}},\ and\ \bibinfo {author} {\bibfnamefont {S.}~\bibnamefont
  {Lambert}},\ }\href {https://doi.org/10.1103/PhysRevD.94.125030} {\bibfield
  {journal} {\bibinfo  {journal} {Phys. Rev. D}\ }\textbf {\bibinfo {volume}
  {94}},\ \bibinfo {pages} {125030} (\bibinfo {year} {2016})}\BibitemShut
  {NoStop}%
\bibitem [{\citenamefont {Long}\ and\ \citenamefont
  {Kosteleck\'y}(2015)}]{PhysRevD.91.092003}%
  \BibitemOpen
  \bibfield  {author} {\bibinfo {author} {\bibfnamefont {J.~C.}\ \bibnamefont
  {Long}}\ and\ \bibinfo {author} {\bibfnamefont {V.~A.}\ \bibnamefont
  {Kosteleck\'y}},\ }\href {https://doi.org/10.1103/PhysRevD.91.092003}
  {\bibfield  {journal} {\bibinfo  {journal} {Phys. Rev. D}\ }\textbf {\bibinfo
  {volume} {91}},\ \bibinfo {pages} {092003} (\bibinfo {year}
  {2015})}\BibitemShut {NoStop}%
\bibitem [{\citenamefont {Shao}\ \emph {et~al.}(2016)\citenamefont {Shao},
  \citenamefont {Tan}, \citenamefont {Tan}, \citenamefont {Yang}, \citenamefont
  {Luo}, \citenamefont {Tobar}, \citenamefont {Bailey}, \citenamefont {Long},
  \citenamefont {Weisman}, \citenamefont {Xu},\ and\ \citenamefont
  {Kosteleck\'y}}]{PhysRevLett.117.071102}%
  \BibitemOpen
  \bibfield  {author} {\bibinfo {author} {\bibfnamefont {C.-G.}\ \bibnamefont
  {Shao}}, \bibinfo {author} {\bibfnamefont {Y.-J.}\ \bibnamefont {Tan}},
  \bibinfo {author} {\bibfnamefont {W.-H.}\ \bibnamefont {Tan}}, \bibinfo
  {author} {\bibfnamefont {S.-Q.}\ \bibnamefont {Yang}}, \bibinfo {author}
  {\bibfnamefont {J.}~\bibnamefont {Luo}}, \bibinfo {author} {\bibfnamefont
  {M.~E.}\ \bibnamefont {Tobar}}, \bibinfo {author} {\bibfnamefont {Q.~G.}\
  \bibnamefont {Bailey}}, \bibinfo {author} {\bibfnamefont {J.~C.}\
  \bibnamefont {Long}}, \bibinfo {author} {\bibfnamefont {E.}~\bibnamefont
  {Weisman}}, \bibinfo {author} {\bibfnamefont {R.}~\bibnamefont {Xu}},\ and\
  \bibinfo {author} {\bibfnamefont {V.~A.}\ \bibnamefont {Kosteleck\'y}},\
  }\href {https://doi.org/10.1103/PhysRevLett.117.071102} {\bibfield  {journal}
  {\bibinfo  {journal} {Phys. Rev. Lett.}\ }\textbf {\bibinfo {volume} {117}},\
  \bibinfo {pages} {071102} (\bibinfo {year} {2016})}\BibitemShut {NoStop}%
\bibitem [{\citenamefont {Shao}\ \emph {et~al.}(2019)\citenamefont {Shao},
  \citenamefont {Chen}, \citenamefont {Tan}, \citenamefont {Yang},
  \citenamefont {Luo}, \citenamefont {Tobar}, \citenamefont {Long},
  \citenamefont {Weisman},\ and\ \citenamefont
  {Kosteleck\'y}}]{PhysRevLett.122.011102}%
  \BibitemOpen
  \bibfield  {author} {\bibinfo {author} {\bibfnamefont {C.-G.}\ \bibnamefont
  {Shao}}, \bibinfo {author} {\bibfnamefont {Y.-F.}\ \bibnamefont {Chen}},
  \bibinfo {author} {\bibfnamefont {Y.-J.}\ \bibnamefont {Tan}}, \bibinfo
  {author} {\bibfnamefont {S.-Q.}\ \bibnamefont {Yang}}, \bibinfo {author}
  {\bibfnamefont {J.}~\bibnamefont {Luo}}, \bibinfo {author} {\bibfnamefont
  {M.~E.}\ \bibnamefont {Tobar}}, \bibinfo {author} {\bibfnamefont {J.~C.}\
  \bibnamefont {Long}}, \bibinfo {author} {\bibfnamefont {E.}~\bibnamefont
  {Weisman}},\ and\ \bibinfo {author} {\bibfnamefont {V.~A.}\ \bibnamefont
  {Kosteleck\'y}},\ }\href {https://doi.org/10.1103/PhysRevLett.122.011102}
  {\bibfield  {journal} {\bibinfo  {journal} {Phys. Rev. Lett.}\ }\textbf
  {\bibinfo {volume} {122}},\ \bibinfo {pages} {011102} (\bibinfo {year}
  {2019})}\BibitemShut {NoStop}%
\bibitem [{\citenamefont {Bailey}\ \emph {et~al.}(2022)\citenamefont {Bailey},
  \citenamefont {James}, \citenamefont {Slone},\ and\ \citenamefont
  {O'Neal-Ault}}]{Bailey:2022wuv}%
  \BibitemOpen
  \bibfield  {author} {\bibinfo {author} {\bibfnamefont {Q.~G.}\ \bibnamefont
  {Bailey}}, \bibinfo {author} {\bibfnamefont {J.~L.}\ \bibnamefont {James}},
  \bibinfo {author} {\bibfnamefont {J.~R.}\ \bibnamefont {Slone}},\ and\
  \bibinfo {author} {\bibfnamefont {K.}~\bibnamefont {O'Neal-Ault}},\
  }\href@noop {} {\  (\bibinfo {year} {2022})},\ \Eprint
  {https://arxiv.org/abs/2210.00605} {arXiv:2210.00605 [gr-qc]} \BibitemShut
  {NoStop}%
\bibitem [{\citenamefont {Shao}\ and\ \citenamefont
  {Bailey}(2018)}]{PhysRevD.98.084049}%
  \BibitemOpen
  \bibfield  {author} {\bibinfo {author} {\bibfnamefont {L.}~\bibnamefont
  {Shao}}\ and\ \bibinfo {author} {\bibfnamefont {Q.~G.}\ \bibnamefont
  {Bailey}},\ }\href {https://doi.org/10.1103/PhysRevD.98.084049} {\bibfield
  {journal} {\bibinfo  {journal} {Phys. Rev. D}\ }\textbf {\bibinfo {volume}
  {98}},\ \bibinfo {pages} {084049} (\bibinfo {year} {2018})}\BibitemShut
  {NoStop}%
\bibitem [{\citenamefont {Shao}(2014)}]{PhysRevLett.112.111103}%
  \BibitemOpen
  \bibfield  {author} {\bibinfo {author} {\bibfnamefont {L.}~\bibnamefont
  {Shao}},\ }\href {https://doi.org/10.1103/PhysRevLett.112.111103} {\bibfield
  {journal} {\bibinfo  {journal} {Phys. Rev. Lett.}\ }\textbf {\bibinfo
  {volume} {112}},\ \bibinfo {pages} {111103} (\bibinfo {year}
  {2014})}\BibitemShut {NoStop}%
\bibitem [{\citenamefont {O'Neal-Ault}\ \emph
  {et~al.}(2021{\natexlab{a}})\citenamefont {O'Neal-Ault}, \citenamefont
  {Bailey}, \citenamefont {Dumerchat}, \citenamefont {Haegel},\ and\
  \citenamefont {Tasson}}]{ONeal-Ault:2021uwu}%
  \BibitemOpen
  \bibfield  {author} {\bibinfo {author} {\bibfnamefont {K.}~\bibnamefont
  {O'Neal-Ault}}, \bibinfo {author} {\bibfnamefont {Q.~G.}\ \bibnamefont
  {Bailey}}, \bibinfo {author} {\bibfnamefont {T.}~\bibnamefont {Dumerchat}},
  \bibinfo {author} {\bibfnamefont {L.}~\bibnamefont {Haegel}},\ and\ \bibinfo
  {author} {\bibfnamefont {J.}~\bibnamefont {Tasson}},\ }\href
  {https://doi.org/10.3390/universe7100380} {\bibfield  {journal} {\bibinfo
  {journal} {Universe}\ }\textbf {\bibinfo {volume} {7}},\ \bibinfo {pages}
  {380} (\bibinfo {year} {2021}{\natexlab{a}})},\ \Eprint
  {https://arxiv.org/abs/2108.06298} {arXiv:2108.06298 [gr-qc]} \BibitemShut
  {NoStop}%
\bibitem [{\citenamefont {Wang}\ \emph {et~al.}(2021)\citenamefont {Wang},
  \citenamefont {Shao},\ and\ \citenamefont {Liu}}]{Wang:2021ctl}%
  \BibitemOpen
  \bibfield  {author} {\bibinfo {author} {\bibfnamefont {Z.}~\bibnamefont
  {Wang}}, \bibinfo {author} {\bibfnamefont {L.}~\bibnamefont {Shao}},\ and\
  \bibinfo {author} {\bibfnamefont {C.}~\bibnamefont {Liu}},\ }\href
  {https://doi.org/10.3847/1538-4357/ac223c} {\bibfield  {journal} {\bibinfo
  {journal} {Astrophys. J.}\ }\textbf {\bibinfo {volume} {921}},\ \bibinfo
  {pages} {158} (\bibinfo {year} {2021})},\ \Eprint
  {https://arxiv.org/abs/2108.02974} {arXiv:2108.02974 [gr-qc]} \BibitemShut
  {NoStop}%
\bibitem [{\citenamefont {Shao}(2020)}]{PhysRevD.101.104019}%
  \BibitemOpen
  \bibfield  {author} {\bibinfo {author} {\bibfnamefont {L.}~\bibnamefont
  {Shao}},\ }\href {https://doi.org/10.1103/PhysRevD.101.104019} {\bibfield
  {journal} {\bibinfo  {journal} {Phys. Rev. D}\ }\textbf {\bibinfo {volume}
  {101}},\ \bibinfo {pages} {104019} (\bibinfo {year} {2020})}\BibitemShut
  {NoStop}%
\bibitem [{\citenamefont {Liu}\ \emph {et~al.}(2020{\natexlab{a}})\citenamefont
  {Liu}, \citenamefont {He}, \citenamefont {Mikulski}, \citenamefont
  {Palenova}, \citenamefont {Williams}, \citenamefont {Creighton},\ and\
  \citenamefont {Tasson}}]{PhysRevD.102.024028}%
  \BibitemOpen
  \bibfield  {author} {\bibinfo {author} {\bibfnamefont {X.}~\bibnamefont
  {Liu}}, \bibinfo {author} {\bibfnamefont {V.~F.}\ \bibnamefont {He}},
  \bibinfo {author} {\bibfnamefont {T.~M.}\ \bibnamefont {Mikulski}}, \bibinfo
  {author} {\bibfnamefont {D.}~\bibnamefont {Palenova}}, \bibinfo {author}
  {\bibfnamefont {C.~E.}\ \bibnamefont {Williams}}, \bibinfo {author}
  {\bibfnamefont {J.}~\bibnamefont {Creighton}},\ and\ \bibinfo {author}
  {\bibfnamefont {J.~D.}\ \bibnamefont {Tasson}},\ }\href
  {https://doi.org/10.1103/PhysRevD.102.024028} {\bibfield  {journal} {\bibinfo
   {journal} {Phys. Rev. D}\ }\textbf {\bibinfo {volume} {102}},\ \bibinfo
  {pages} {024028} (\bibinfo {year} {2020}{\natexlab{a}})}\BibitemShut
  {NoStop}%
\bibitem [{\citenamefont {Bonder}\ and\ \citenamefont
  {Leon}(2017)}]{Bonder:2017dpb}%
  \BibitemOpen
  \bibfield  {author} {\bibinfo {author} {\bibfnamefont {Y.}~\bibnamefont
  {Bonder}}\ and\ \bibinfo {author} {\bibfnamefont {G.}~\bibnamefont {Leon}},\
  }\href {https://doi.org/10.1103/PhysRevD.96.044036} {\bibfield  {journal}
  {\bibinfo  {journal} {Phys. Rev. D}\ }\textbf {\bibinfo {volume} {96}},\
  \bibinfo {pages} {044036} (\bibinfo {year} {2017})},\ \Eprint
  {https://arxiv.org/abs/1704.05894} {arXiv:1704.05894 [gr-qc]} \BibitemShut
  {NoStop}%
\bibitem [{\citenamefont {Bonder}\ and\ \citenamefont
  {Peterson}(2020)}]{Bonder:2020fpn}%
  \BibitemOpen
  \bibfield  {author} {\bibinfo {author} {\bibfnamefont {Y.}~\bibnamefont
  {Bonder}}\ and\ \bibinfo {author} {\bibfnamefont {C.}~\bibnamefont
  {Peterson}},\ }\href {https://doi.org/10.1103/PhysRevD.101.064056} {\bibfield
   {journal} {\bibinfo  {journal} {Phys. Rev. D}\ }\textbf {\bibinfo {volume}
  {101}},\ \bibinfo {pages} {064056} (\bibinfo {year} {2020})},\ \Eprint
  {https://arxiv.org/abs/2001.09217} {arXiv:2001.09217 [gr-qc]} \BibitemShut
  {NoStop}%
\bibitem [{\citenamefont {O'Neal-Ault}\ \emph
  {et~al.}(2021{\natexlab{b}})\citenamefont {O'Neal-Ault}, \citenamefont
  {Bailey},\ and\ \citenamefont {Nilsson}}]{ONeal-Ault:2020ebv}%
  \BibitemOpen
  \bibfield  {author} {\bibinfo {author} {\bibfnamefont {K.}~\bibnamefont
  {O'Neal-Ault}}, \bibinfo {author} {\bibfnamefont {Q.~G.}\ \bibnamefont
  {Bailey}},\ and\ \bibinfo {author} {\bibfnamefont {N.~A.}\ \bibnamefont
  {Nilsson}},\ }\href {https://doi.org/10.1103/PhysRevD.103.044010} {\bibfield
  {journal} {\bibinfo  {journal} {Phys. Rev. D}\ }\textbf {\bibinfo {volume}
  {103}},\ \bibinfo {pages} {044010} (\bibinfo {year} {2021}{\natexlab{b}})},\
  \Eprint {https://arxiv.org/abs/2009.00949} {arXiv:2009.00949 [gr-qc]}
  \BibitemShut {NoStop}%
\bibitem [{\citenamefont {Bailey}(2016)}]{Bailey:2016ezm}%
  \BibitemOpen
  \bibfield  {author} {\bibinfo {author} {\bibfnamefont {Q.~G.}\ \bibnamefont
  {Bailey}},\ }\href {https://doi.org/10.1103/PhysRevD.94.065029} {\bibfield
  {journal} {\bibinfo  {journal} {Phys. Rev. D}\ }\textbf {\bibinfo {volume}
  {94}},\ \bibinfo {pages} {065029} (\bibinfo {year} {2016})},\ \Eprint
  {https://arxiv.org/abs/1608.00267} {arXiv:1608.00267 [gr-qc]} \BibitemShut
  {NoStop}%
\bibitem [{\citenamefont {Bonder}(2015)}]{Bonder:2015maa}%
  \BibitemOpen
  \bibfield  {author} {\bibinfo {author} {\bibfnamefont {Y.}~\bibnamefont
  {Bonder}},\ }\href {https://doi.org/10.1103/PhysRevD.91.125002} {\bibfield
  {journal} {\bibinfo  {journal} {Phys. Rev. D}\ }\textbf {\bibinfo {volume}
  {91}},\ \bibinfo {pages} {125002} (\bibinfo {year} {2015})},\ \Eprint
  {https://arxiv.org/abs/1504.03636} {arXiv:1504.03636 [gr-qc]} \BibitemShut
  {NoStop}%
\bibitem [{\citenamefont {Bailey}(2020)}]{Bailey:2019rjj}%
  \BibitemOpen
  \bibfield  {author} {\bibinfo {author} {\bibfnamefont {Q.~G.}\ \bibnamefont
  {Bailey}},\ }in\ \href {https://doi.org/10.1142/9789811213984_0018} {\emph
  {\bibinfo {booktitle} {{8th Meeting on CPT and Lorentz Symmetry}}}}\
  (\bibinfo {year} {2020})\ pp.\ \bibinfo {pages} {69--72},\ \Eprint
  {https://arxiv.org/abs/1906.08657} {arXiv:1906.08657 [gr-qc]} \BibitemShut
  {NoStop}%
\bibitem [{\citenamefont {Reyes}\ and\ \citenamefont
  {Schreck}(2021)}]{Reyes:2021cpx}%
  \BibitemOpen
  \bibfield  {author} {\bibinfo {author} {\bibfnamefont {C.~M.}\ \bibnamefont
  {Reyes}}\ and\ \bibinfo {author} {\bibfnamefont {M.}~\bibnamefont
  {Schreck}},\ }\href {https://doi.org/10.1103/PhysRevD.104.124042} {\bibfield
  {journal} {\bibinfo  {journal} {Phys. Rev. D}\ }\textbf {\bibinfo {volume}
  {104}},\ \bibinfo {pages} {124042} (\bibinfo {year} {2021})},\ \Eprint
  {https://arxiv.org/abs/2105.05954} {arXiv:2105.05954 [gr-qc]} \BibitemShut
  {NoStop}%
\bibitem [{\citenamefont {Kostelecky}\ and\ \citenamefont
  {Russell}(2008)}]{Kostelecky:2008ts}%
  \BibitemOpen
  \bibfield  {author} {\bibinfo {author} {\bibfnamefont {V.~A.}\ \bibnamefont
  {Kostelecky}}\ and\ \bibinfo {author} {\bibfnamefont {N.}~\bibnamefont
  {Russell}},\ }\href@noop {} {\  (\bibinfo {year} {2008})},\ \Eprint
  {https://arxiv.org/abs/0801.0287} {arXiv:0801.0287 [hep-ph]} \BibitemShut
  {NoStop}%
\bibitem [{\citenamefont {Bluhm}\ \emph {et~al.}(2019)\citenamefont {Bluhm},
  \citenamefont {Bossi},\ and\ \citenamefont {Wen}}]{Bluhm:2019ato}%
  \BibitemOpen
  \bibfield  {author} {\bibinfo {author} {\bibfnamefont {R.}~\bibnamefont
  {Bluhm}}, \bibinfo {author} {\bibfnamefont {H.}~\bibnamefont {Bossi}},\ and\
  \bibinfo {author} {\bibfnamefont {Y.}~\bibnamefont {Wen}},\ }\href
  {https://doi.org/10.1103/PhysRevD.100.084022} {\bibfield  {journal} {\bibinfo
   {journal} {Phys. Rev. D}\ }\textbf {\bibinfo {volume} {100}},\ \bibinfo
  {pages} {084022} (\bibinfo {year} {2019})},\ \Eprint
  {https://arxiv.org/abs/1907.13209} {arXiv:1907.13209 [gr-qc]} \BibitemShut
  {NoStop}%
\bibitem [{\citenamefont {Bluhm}\ and\ \citenamefont
  {Yang}(2021)}]{Bluhm:2021lzf}%
  \BibitemOpen
  \bibfield  {author} {\bibinfo {author} {\bibfnamefont {R.}~\bibnamefont
  {Bluhm}}\ and\ \bibinfo {author} {\bibfnamefont {Y.}~\bibnamefont {Yang}},\
  }\href {https://doi.org/10.3390/sym13040660} {\bibfield  {journal} {\bibinfo
  {journal} {Symmetry}\ }\textbf {\bibinfo {volume} {13}},\ \bibinfo {pages}
  {660} (\bibinfo {year} {2021})},\ \Eprint {https://arxiv.org/abs/2104.05879}
  {arXiv:2104.05879 [gr-qc]} \BibitemShut {NoStop}%
\bibitem [{\citenamefont {Arnowitt}\ \emph {et~al.}(2008)\citenamefont
  {Arnowitt}, \citenamefont {Deser},\ and\ \citenamefont
  {Misner}}]{Arnowitt:1962hi}%
  \BibitemOpen
  \bibfield  {author} {\bibinfo {author} {\bibfnamefont {R.~L.}\ \bibnamefont
  {Arnowitt}}, \bibinfo {author} {\bibfnamefont {S.}~\bibnamefont {Deser}},\
  and\ \bibinfo {author} {\bibfnamefont {C.~W.}\ \bibnamefont {Misner}},\
  }\href {https://doi.org/10.1007/s10714-008-0661-1} {\bibfield  {journal}
  {\bibinfo  {journal} {Gen. Rel. Grav.}\ }\textbf {\bibinfo {volume} {40}},\
  \bibinfo {pages} {1997} (\bibinfo {year} {2008})},\ \Eprint
  {https://arxiv.org/abs/gr-qc/0405109} {arXiv:gr-qc/0405109} \BibitemShut
  {NoStop}%
\bibitem [{\citenamefont {Zhu}\ \emph {et~al.}(2022)\citenamefont {Zhu},
  \citenamefont {Zhao},\ and\ \citenamefont {Wang}}]{wang}%
  \BibitemOpen
  \bibfield  {author} {\bibinfo {author} {\bibfnamefont {T.}~\bibnamefont
  {Zhu}}, \bibinfo {author} {\bibfnamefont {W.}~\bibnamefont {Zhao}},\ and\
  \bibinfo {author} {\bibfnamefont {A.}~\bibnamefont {Wang}},\ }\href
  {https://doi.org/10.48550/ARXIV.2210.05259} {\bibinfo {title} {Polarized
  primordial gravitational waves in spatial covariant gravities}} (\bibinfo
  {year} {2022})\BibitemShut {NoStop}%
\bibitem [{\citenamefont {Pettorino}\ and\ \citenamefont
  {Amendola}(2015)}]{Pettorino:2014bka}%
  \BibitemOpen
  \bibfield  {author} {\bibinfo {author} {\bibfnamefont {V.}~\bibnamefont
  {Pettorino}}\ and\ \bibinfo {author} {\bibfnamefont {L.}~\bibnamefont
  {Amendola}},\ }\href {https://doi.org/10.1016/j.physletb.2015.02.007}
  {\bibfield  {journal} {\bibinfo  {journal} {Phys. Lett. B}\ }\textbf
  {\bibinfo {volume} {742}},\ \bibinfo {pages} {353} (\bibinfo {year}
  {2015})},\ \Eprint {https://arxiv.org/abs/1408.2224} {arXiv:1408.2224
  [astro-ph.CO]} \BibitemShut {NoStop}%
\bibitem [{\citenamefont {Nishizawa}(2018)}]{Nishizawa:2017nef}%
  \BibitemOpen
  \bibfield  {author} {\bibinfo {author} {\bibfnamefont {A.}~\bibnamefont
  {Nishizawa}},\ }\href {https://doi.org/10.1103/PhysRevD.97.104037} {\bibfield
   {journal} {\bibinfo  {journal} {Phys. Rev. D}\ }\textbf {\bibinfo {volume}
  {97}},\ \bibinfo {pages} {104037} (\bibinfo {year} {2018})},\ \Eprint
  {https://arxiv.org/abs/1710.04825} {arXiv:1710.04825 [gr-qc]} \BibitemShut
  {NoStop}%
\bibitem [{\citenamefont {Baker}\ \emph {et~al.}(2022)\citenamefont {Baker}
  \emph {et~al.}}]{Baker:2022rhh}%
  \BibitemOpen
  \bibfield  {author} {\bibinfo {author} {\bibfnamefont {T.}~\bibnamefont
  {Baker}} \emph {et~al.},\ }\href@noop {} {\  (\bibinfo {year} {2022})},\
  \Eprint {https://arxiv.org/abs/2203.00566} {arXiv:2203.00566 [gr-qc]}
  \BibitemShut {NoStop}%
\bibitem [{\citenamefont {Robbers}\ \emph {et~al.}(2008)\citenamefont
  {Robbers}, \citenamefont {Afshordi},\ and\ \citenamefont
  {Doran}}]{Robbers:2007ca}%
  \BibitemOpen
  \bibfield  {author} {\bibinfo {author} {\bibfnamefont {G.}~\bibnamefont
  {Robbers}}, \bibinfo {author} {\bibfnamefont {N.}~\bibnamefont {Afshordi}},\
  and\ \bibinfo {author} {\bibfnamefont {M.}~\bibnamefont {Doran}},\ }\href
  {https://doi.org/10.1103/PhysRevLett.100.111101} {\bibfield  {journal}
  {\bibinfo  {journal} {Phys. Rev. Lett.}\ }\textbf {\bibinfo {volume} {100}},\
  \bibinfo {pages} {111101} (\bibinfo {year} {2008})},\ \Eprint
  {https://arxiv.org/abs/0708.3235} {arXiv:0708.3235 [astro-ph]} \BibitemShut
  {NoStop}%
\bibitem [{\citenamefont {Lagos}\ \emph {et~al.}(2019)\citenamefont {Lagos},
  \citenamefont {Fishbach}, \citenamefont {Landry},\ and\ \citenamefont
  {Holz}}]{Lagos:2019kds}%
  \BibitemOpen
  \bibfield  {author} {\bibinfo {author} {\bibfnamefont {M.}~\bibnamefont
  {Lagos}}, \bibinfo {author} {\bibfnamefont {M.}~\bibnamefont {Fishbach}},
  \bibinfo {author} {\bibfnamefont {P.}~\bibnamefont {Landry}},\ and\ \bibinfo
  {author} {\bibfnamefont {D.~E.}\ \bibnamefont {Holz}},\ }\href
  {https://doi.org/10.1103/PhysRevD.99.083504} {\bibfield  {journal} {\bibinfo
  {journal} {Phys. Rev. D}\ }\textbf {\bibinfo {volume} {99}},\ \bibinfo
  {pages} {083504} (\bibinfo {year} {2019})},\ \Eprint
  {https://arxiv.org/abs/1901.03321} {arXiv:1901.03321 [astro-ph.CO]}
  \BibitemShut {NoStop}%
\bibitem [{\citenamefont {Kostelecky}\ and\ \citenamefont
  {Tasson}(2009)}]{Kostelecky:2008in}%
  \BibitemOpen
  \bibfield  {author} {\bibinfo {author} {\bibfnamefont {V.~A.}\ \bibnamefont
  {Kostelecky}}\ and\ \bibinfo {author} {\bibfnamefont {J.}~\bibnamefont
  {Tasson}},\ }\href {https://doi.org/10.1103/PhysRevLett.102.010402}
  {\bibfield  {journal} {\bibinfo  {journal} {Phys. Rev. Lett.}\ }\textbf
  {\bibinfo {volume} {102}},\ \bibinfo {pages} {010402} (\bibinfo {year}
  {2009})},\ \Eprint {https://arxiv.org/abs/0810.1459} {arXiv:0810.1459
  [gr-qc]} \BibitemShut {NoStop}%
\bibitem [{\citenamefont {Pihan-Le~Bars}\ \emph {et~al.}(2019)\citenamefont
  {Pihan-Le~Bars} \emph {et~al.}}]{Pihan-LeBars:2019qsd}%
  \BibitemOpen
  \bibfield  {author} {\bibinfo {author} {\bibfnamefont {H.}~\bibnamefont
  {Pihan-Le~Bars}} \emph {et~al.},\ }\href
  {https://doi.org/10.1103/PhysRevLett.123.231102} {\bibfield  {journal}
  {\bibinfo  {journal} {Phys. Rev. Lett.}\ }\textbf {\bibinfo {volume} {123}},\
  \bibinfo {pages} {231102} (\bibinfo {year} {2019})},\ \Eprint
  {https://arxiv.org/abs/1912.03030} {arXiv:1912.03030 [physics.space-ph]}
  \BibitemShut {NoStop}%
\bibitem [{\citenamefont {Liu}\ \emph {et~al.}(2020{\natexlab{b}})\citenamefont
  {Liu}, \citenamefont {He}, \citenamefont {Mikulski}, \citenamefont
  {Palenova}, \citenamefont {Williams}, \citenamefont {Creighton},\ and\
  \citenamefont {Tasson}}]{Liu:2020slm}%
  \BibitemOpen
  \bibfield  {author} {\bibinfo {author} {\bibfnamefont {X.}~\bibnamefont
  {Liu}}, \bibinfo {author} {\bibfnamefont {V.~F.}\ \bibnamefont {He}},
  \bibinfo {author} {\bibfnamefont {T.~M.}\ \bibnamefont {Mikulski}}, \bibinfo
  {author} {\bibfnamefont {D.}~\bibnamefont {Palenova}}, \bibinfo {author}
  {\bibfnamefont {C.~E.}\ \bibnamefont {Williams}}, \bibinfo {author}
  {\bibfnamefont {J.}~\bibnamefont {Creighton}},\ and\ \bibinfo {author}
  {\bibfnamefont {J.~D.}\ \bibnamefont {Tasson}},\ }\href
  {https://doi.org/10.1103/PhysRevD.102.024028} {\bibfield  {journal} {\bibinfo
   {journal} {Phys. Rev. D}\ }\textbf {\bibinfo {volume} {102}},\ \bibinfo
  {pages} {024028} (\bibinfo {year} {2020}{\natexlab{b}})},\ \Eprint
  {https://arxiv.org/abs/2005.03121} {arXiv:2005.03121 [gr-qc]} \BibitemShut
  {NoStop}%
\bibitem [{\citenamefont {Haegel}\ \emph {et~al.}(2022)\citenamefont {Haegel},
  \citenamefont {O'Neal-Ault}, \citenamefont {Bailey}, \citenamefont {Tasson},
  \citenamefont {Bloom},\ and\ \citenamefont {Shao}}]{Haegel:2022ymk}%
  \BibitemOpen
  \bibfield  {author} {\bibinfo {author} {\bibfnamefont {L.}~\bibnamefont
  {Haegel}}, \bibinfo {author} {\bibfnamefont {K.}~\bibnamefont {O'Neal-Ault}},
  \bibinfo {author} {\bibfnamefont {Q.~G.}\ \bibnamefont {Bailey}}, \bibinfo
  {author} {\bibfnamefont {J.~D.}\ \bibnamefont {Tasson}}, \bibinfo {author}
  {\bibfnamefont {M.}~\bibnamefont {Bloom}},\ and\ \bibinfo {author}
  {\bibfnamefont {L.}~\bibnamefont {Shao}},\ }\href@noop {} {\  (\bibinfo
  {year} {2022})},\ \Eprint {https://arxiv.org/abs/2210.04481}
  {arXiv:2210.04481 [gr-qc]} \BibitemShut {NoStop}%
\bibitem [{\citenamefont {Boyle}\ and\ \citenamefont
  {Steinhardt}(2008)}]{Boyle:2005se}%
  \BibitemOpen
  \bibfield  {author} {\bibinfo {author} {\bibfnamefont {L.~A.}\ \bibnamefont
  {Boyle}}\ and\ \bibinfo {author} {\bibfnamefont {P.~J.}\ \bibnamefont
  {Steinhardt}},\ }\href {https://doi.org/10.1103/PhysRevD.77.063504}
  {\bibfield  {journal} {\bibinfo  {journal} {Phys. Rev. D}\ }\textbf {\bibinfo
  {volume} {77}},\ \bibinfo {pages} {063504} (\bibinfo {year} {2008})},\
  \Eprint {https://arxiv.org/abs/astro-ph/0512014} {arXiv:astro-ph/0512014}
  \BibitemShut {NoStop}%
\bibitem [{\citenamefont {Watanabe}\ and\ \citenamefont
  {Komatsu}(2006)}]{Watanabe:2006qe}%
  \BibitemOpen
  \bibfield  {author} {\bibinfo {author} {\bibfnamefont {Y.}~\bibnamefont
  {Watanabe}}\ and\ \bibinfo {author} {\bibfnamefont {E.}~\bibnamefont
  {Komatsu}},\ }\href {https://doi.org/10.1103/PhysRevD.73.123515} {\bibfield
  {journal} {\bibinfo  {journal} {Phys. Rev. D}\ }\textbf {\bibinfo {volume}
  {73}},\ \bibinfo {pages} {123515} (\bibinfo {year} {2006})},\ \Eprint
  {https://arxiv.org/abs/astro-ph/0604176} {arXiv:astro-ph/0604176}
  \BibitemShut {NoStop}%
\bibitem [{\citenamefont {Kostelecky}\ and\ \citenamefont
  {Tasson}(2011)}]{Kostelecky:2010ze}%
  \BibitemOpen
  \bibfield  {author} {\bibinfo {author} {\bibfnamefont {A.~V.}\ \bibnamefont
  {Kostelecky}}\ and\ \bibinfo {author} {\bibfnamefont {J.~D.}\ \bibnamefont
  {Tasson}},\ }\href {https://doi.org/10.1103/PhysRevD.83.016013} {\bibfield
  {journal} {\bibinfo  {journal} {Phys. Rev. D}\ }\textbf {\bibinfo {volume}
  {83}},\ \bibinfo {pages} {016013} (\bibinfo {year} {2011})},\ \Eprint
  {https://arxiv.org/abs/1006.4106} {arXiv:1006.4106 [gr-qc]} \BibitemShut
  {NoStop}%
\bibitem [{\citenamefont {Howlett}(1966)}]{howlett_1966}%
  \BibitemOpen
  \bibfield  {author} {\bibinfo {author} {\bibfnamefont {J.}~\bibnamefont
  {Howlett}},\ }\href {https://doi.org/10.2307/3614753} {\bibfield  {journal}
  {\bibinfo  {journal} {The Mathematical Gazette}\ }\textbf {\bibinfo {volume}
  {50}},\ \bibinfo {pages} {358–359} (\bibinfo {year} {1966})}\BibitemShut
  {NoStop}%
\end{thebibliography}%

\end{document}